\documentclass{aastex62}

\usepackage{graphicx}
\usepackage{graphics}
\usepackage{amssymb}
\usepackage{bm}
\usepackage{times}

\newcommand{\fhi}{f_{\rm HI}}
\newcommand{\beq}{\begin{equation}}
\newcommand{\eeq}{\end{equation}}

\newcommand{\msun}{\textrm{M}_\odot}

\newcommand{\mstar}{\rm{M_{\star}}}

\newcommand{\kms}{km~s$^{-1}$}
\newcommand{\mhi}{\rm M_{HI}}
\newcommand{\mb}{\rm M_B}

\newcommand{\hi}{H{\sc i}}

\newcommand{\lya}{Lyman-$\alpha$}

\newcommand{\hii}{H{\sc i}\,21cm}

\shorttitle{The H{\sc i} Mass of Green Pea Galaxies}
\shortauthors{Kanekar et al.}
\begin{document}
\title{The Atomic Gas Mass of Green Pea Galaxies}

\correspondingauthor{Nissim Kanekar}
\email{nkanekar@ncra.tifr.res.in}

\author{N. Kanekar} 
\affiliation{National Centre for Radio Astrophysics, Tata Institute of Fundamental Research, 
Pune University, Pune 411007, India}

\author{T. Ghosh}
\affiliation{Green Bank Observatory, P.O. Box 2, Green Bank, WV 24944, USA}

\author{J. Rhoads}
\affiliation{Astrophysics Division, NASA Goddard Space Flight Center, Greenbelt, MD 20771, USA}
\affiliation{School of Earth and Space Exploration, Arizona State University, Tempe, AZ 85287, USA}

\author{S. Malhotra}
\affiliation{Astrophysics Division, NASA Goddard Space Flight Center, Greenbelt, MD 20771, USA}
\affiliation{School of Earth and Space Exploration, Arizona State University, Tempe, AZ 85287, USA}

\author{S. Harish}
\affiliation{School of Earth and Space Exploration, Arizona State University, Tempe, AZ 85287, USA}

\author{J. N. Chengalur}
\affiliation{National Centre for Radio Astrophysics, Tata Institute of Fundamental Research, 
Pune University, Pune 411007, India}

\author{K. M. Jones}
\affiliation{Department of Physics and Astronomy, University of Kansas, 1082 Malott, 1251 Wescoe Hall Dr.
Lawrence, KS 66045 }

\begin{abstract}
We have used the Arecibo Telescope and the Green Bank Telescope to carry out a deep search for 
H{\sc i}~21\,cm emission from a large sample of Green Pea galaxies, yielding 19 detections, and 
21 upper limits on the H{\sc i} mass. We obtain H{\sc i} masses of 
$\rm M_{HI} \approx (4-300) \times 10^8 \, \rm M_\odot$ for the detections, with a median H{\sc i} 
mass of $\approx 2.6 \times 10^9 \, \rm M_\odot$; for the non-detections, the median $3\sigma$ upper 
limit on the H{\sc i} mass is $\approx 5.5 \times 10^8 \, \rm M_\odot$. These are the first estimates of the 
atomic gas content of Green Pea galaxies. We find that the H{\sc i}-to-stellar mass ratio in Green Peas 
is consistent with trends identified in star-forming galaxies in the local Universe. However, the median H{\sc i} 
depletion timescale in Green Peas is $\approx 0.6$~Gyr, an order of magnitude lower than that obtained 
in local star-forming galaxies. This implies that Green Peas consume their atomic gas on very short 
timescales. A significant fraction of the Green Peas of our sample lie $\gtrsim 0.6$~dex ($2\sigma$) 
above the local $\rm M_{HI} - M_B$ relation, suggesting recent gas accretion. Further, $\approx 30$\% of 
the Green Peas are more than $\pm 2\sigma$ deviant from this relation, suggesting possible bimodality 
in the Green Pea population. We obtain a low H{\sc i}~21\,cm detection rate in the Green Peas with the 
highest O32~$\equiv$~[O{\sc iii}]$\lambda$5007/[O{\sc ii}]$\lambda$3727 luminosity ratios, O32~$> 10$, 
consistent with the high expected Lyman-continuum leakage from these galaxies.

\end{abstract}


\section*{}
\noindent This paper is dedicated to the Arecibo Observatory and its people.\\
{\it De estas calles que ahondan el poniente, \\ Una habr{\'a} (no s{\'e} cual) que he recorrido, \\ Ya por {\'u}ltima vez, ...\footnote{``L{\'i}mites'', Jorge Lu{\'i}s Borges.}}

\vskip 0.15in

\section{Introduction} 
\label{sec:intro}

The nature of ``Green Pea'' galaxies, the low-redshift ($z \lesssim 0.3$) extreme emission-line 
galaxies identified by the Galaxy Zoo project \citep{cardamone09}, has been of much 
interest over the last decade. Their low metallicity and dust content, strong nebular lines, 
compact or interacting morphology, and intense star formation activity are all reminiscent of 
high-$z$ \lya\ emitters \citep[e.g.][]{izotov11,yang17c,jiang19}. Indeed, for Green Peas 
studied at ultraviolet (UV) wavelengths, the \lya\ equivalent width distribution is similar to that of \lya\ 
emitters at $z \gtrsim 2.8$ \citep{yang16}, while the \lya\ and UV continuum sizes are similar to 
those of \lya\ emitters at $z \approx 3-6$ \citep{yang17c}. Green Peas show a high [O{\sc iii}]$\lambda$5007/[O{\sc ii}]$\lambda$3727
luminosity ratio, similar to many high-$z$ star-forming galaxies, indicating optically-thin ionized regions 
\citep[e.g.][]{jaskot13,nakajima20}. Perhaps most interesting, and unlike most galaxies in the 
low-$z$ Universe, Green Peas have been found to commonly show leakage of Lyman-continuum radiation, 
with escape fractions of $\approx 2.5-73$\% \citep{izotov16,izotov18b,izotov18a}. Such Lyman-continuum 
radiation escaping from star-forming galaxies is expected to have been the prime cause of the 
reionization of the Universe, at $z \gtrsim 6$ \citep[e.g.][]{fan06}; however, 
the dependence of the escape fraction on local conditions is still not understood today. Galaxies 
like the Green Peas that show strong Lyman-continuum leakage are the best low-$z$ analogs of the 
galaxies that drove cosmological reionization, and offer the exciting possibility of understanding 
this critical process in the nearby Universe.

While detailed optical and UV imaging and spectroscopic studies have characterized the stellar, 
nebular and star-formation properties of the Green Peas 
\citep[e.g.][]{amorin10,izotov11,izotov18a,jaskot14,yang16,yang17c,lofthouse17,jiang19}, little is known 
about the primary fuel for star-formation in these galaxies, the neutral atomic or molecular gas.
As such, the cause of the intense starburst activity in the Green Peas remains unclear.
Further, there is a natural tension between requiring cold neutral gas to fuel the starburst activity 
and having a sufficiently low \hi\ column density to allow the resonantly scattered 
\lya\ and Lyman continuum to escape. This suggests that the \hi\ column density distribution in 
Green Peas may be highly non-uniform, with \hi\ porosity playing a key role \citep[but see][]{henry15}.


At present, only two Green Peas have published searches for \hii\ emission, both yielding upper limits 
on the \hi\ mass of the galaxy \citep{pardy14,mckinney19}. We report here Arecibo Telescope (hereafter, 
Arecibo) and Green Bank Telescope (GBT) \hii\ spectroscopy of a large sample of Green Peas at $z \approx 0.02-0.1$, 
which allow us to measure the atomic gas mass of these galaxies for the first time.\footnote{We assume a 
flat $\Lambda$-Cold Dark Matter cosmology, with $\Omega_\Lambda = 0.685$, $\Omega_m = 0.315$, 
H$_0 = 67.4$~\kms~Mpc$^{-1}$ \citep{planck20}.}

\section{Observations, Data analysis, and Results}

\citet{jiang19} have compiled the most comprehensive Green Pea galaxy sample to date, consisting of 
approximately 1000 galaxies at $0.01 \lesssim z \lesssim 0.41$, identified from the Sloan Digital Sky 
Survey spectroscopic Data Release~13. We used the correlation between B-band luminosity and \hi\ mass 
\citep[e.g.][]{denes14} to pre-select Green Peas from the above sample with \hi\ masses high enough 
to show detectable \hii\ emission with Arecibo and the GBT in reasonable integration time (few hours). 
Our targets span a wide range of absolute B-band magnitudes ($-20.0 \leq \mb \leq -16.1$) and gas-phase metallicities 
($7.6 \leq$~12+[O/H]~$\leq 8.35$). We also carried out two-sample Kolmogorov-Smirnov tests to compare the 
distributions of metallicity, stellar mass, and absolute B-magnitude in our target sample with those of the parent 
Green Pea sample of \citet{jiang19}. We find that the data are consistent with the null hypothesis that the two samples 
are drawn from the same distribution, in all three parameters.

We used Arecibo and the GBT to carry out a search for \hii\ emission from 44 Green Peas, at 
$z \approx 0.02 - 0.1$ (proposals GBT/19A-301: PI Malhotra; Arecibo/A3302: PI Rhoads), between 
February and August~2019. To use the complementary strengths of Arecibo and the GBT, we observed 
lower-redshift targets ($z \lesssim 0.05$) with higher expected \hii\ line flux densities over the 
entire northern and equatorial sky using the GBT. With Arecibo, we broadened the selection to include 
Green Peas with lower expected \hii\ line flux densities and higher redshifts ($z \lesssim 0.1$),
within the region of sky accessible to the telescope.

The Arecibo observations used the L-wide receiver, the WAPP backend, two orthogonal polarizations, and a 
25~MHz band sub-divided into 
4096 spectral channels and centered on the redshifted \hii\ line frequency. The GBT observations used the 
L-band receiver with the VEGAS spectrometer as the backend, two polarizations, and a 23.44~MHz bandwidth 
sub-divided into 8192 channels, and centered on the redshifted \hii\ line frequency. Position-switching, 
with 5m On and Off scans, was used to calibrate the system bandpass, while the system temperatures were 
measured using a blinking noise diode at the GBT, and a separate noise diode, switched on and off for 10~sec,
at Arecibo. Online doppler tracking was not used. The total time on each source 
ranged from $0.75-4.5$~hours, depending on the galaxy redshift, RFI conditions, and observing exigencies.

All data were analysed in the IDL package, following standard procedures, with the package {\sc gbtidl} used 
for the GBT data. Each On/Off pair was initially calibrated and the final spectrum, for each polarization, 
shifted into the barycentric frame. Each spectrum was then inspected for the presence of radio frequency 
interference (RFI) or systematic effects in the spectral baseline; spectra showing non-gaussian behaviour 
within $\approx \pm 200$~\kms\ of the expected redshifted \hii\ line frequency were removed from the analysis. 
For each source, the remaining spectra, from both polarizations, were median-averaged together, with the median 
used to obtain a more conservative (i.e. less sensitive to outliers) estimate of the average. For four sources, 
two from each telescope, all spectra were
affected by RFI around the expected redshifted line frequency, and the data were essentially unuseable.

\begin{figure*}[t!]
\centering
\includegraphics[width=1.7in]{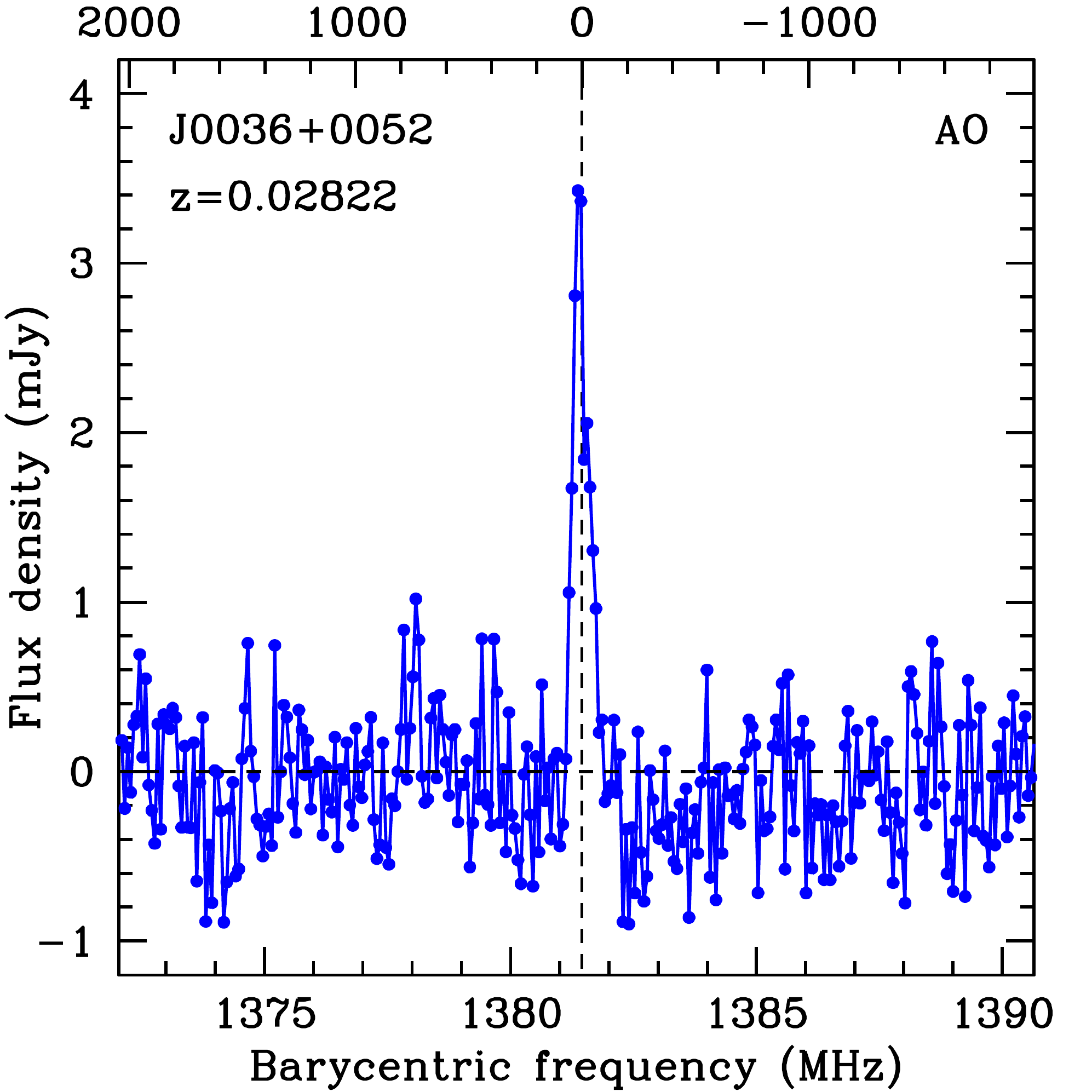}
\includegraphics[width=1.7in]{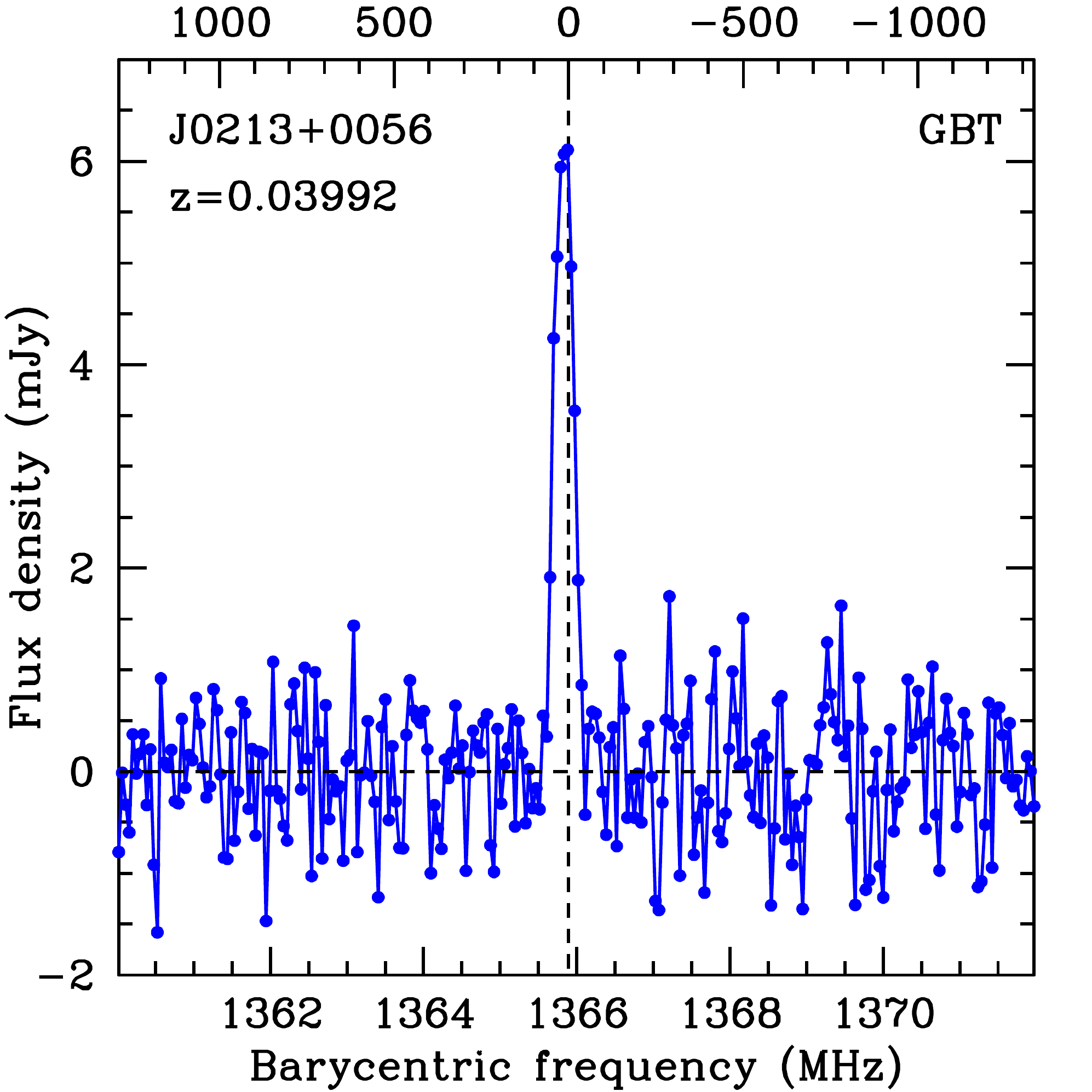}
\includegraphics[width=1.7in]{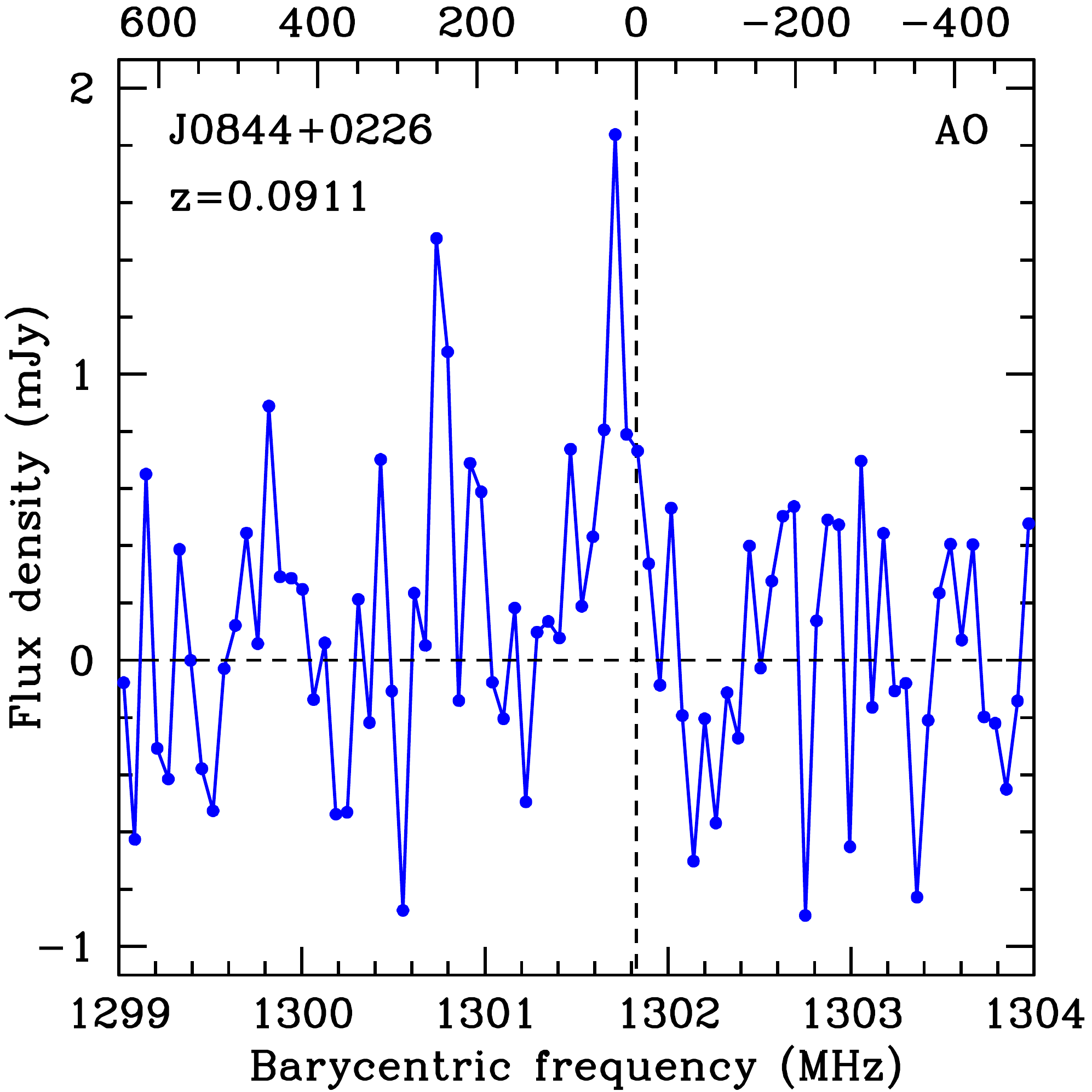}
\includegraphics[width=1.7in]{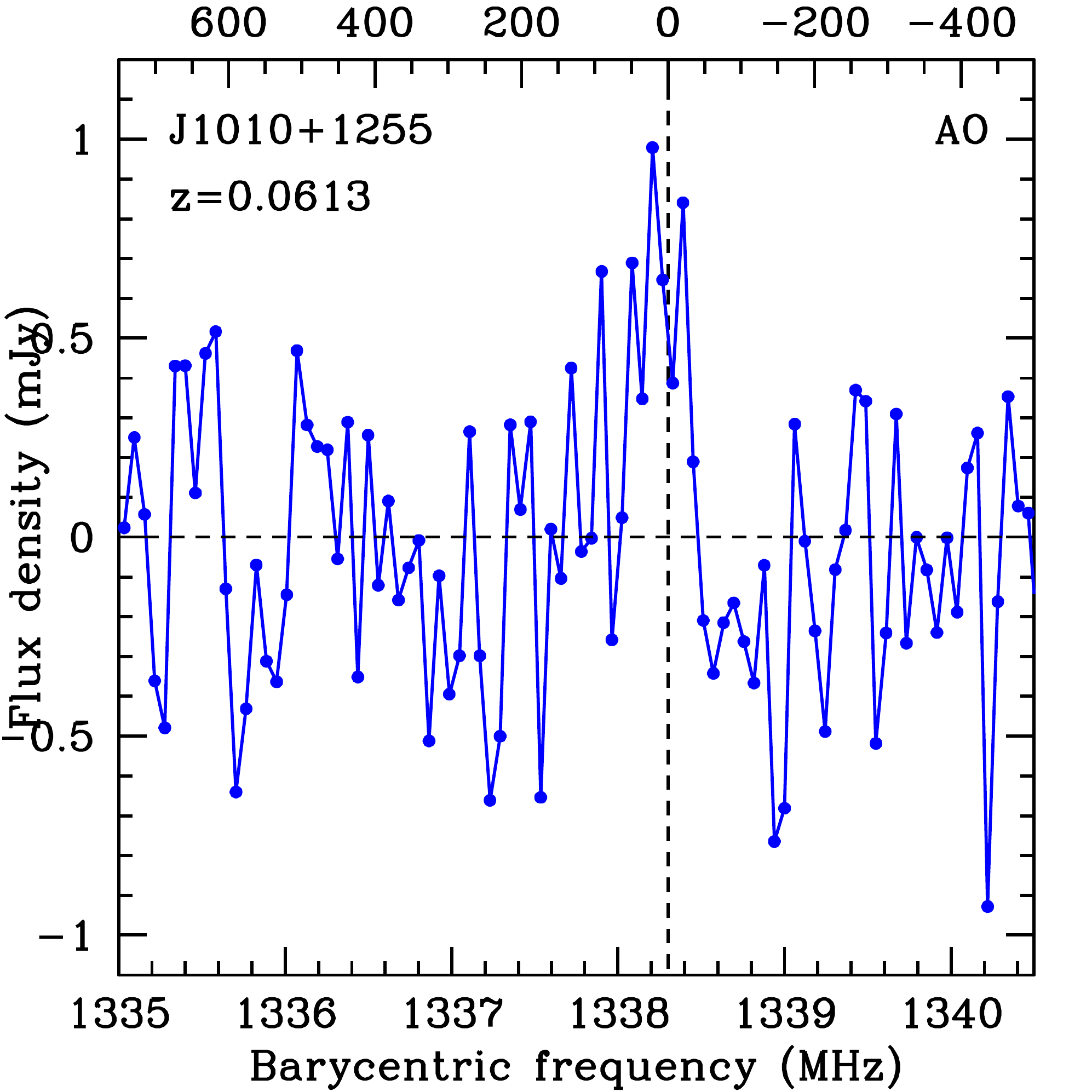}
\includegraphics[width=1.7in]{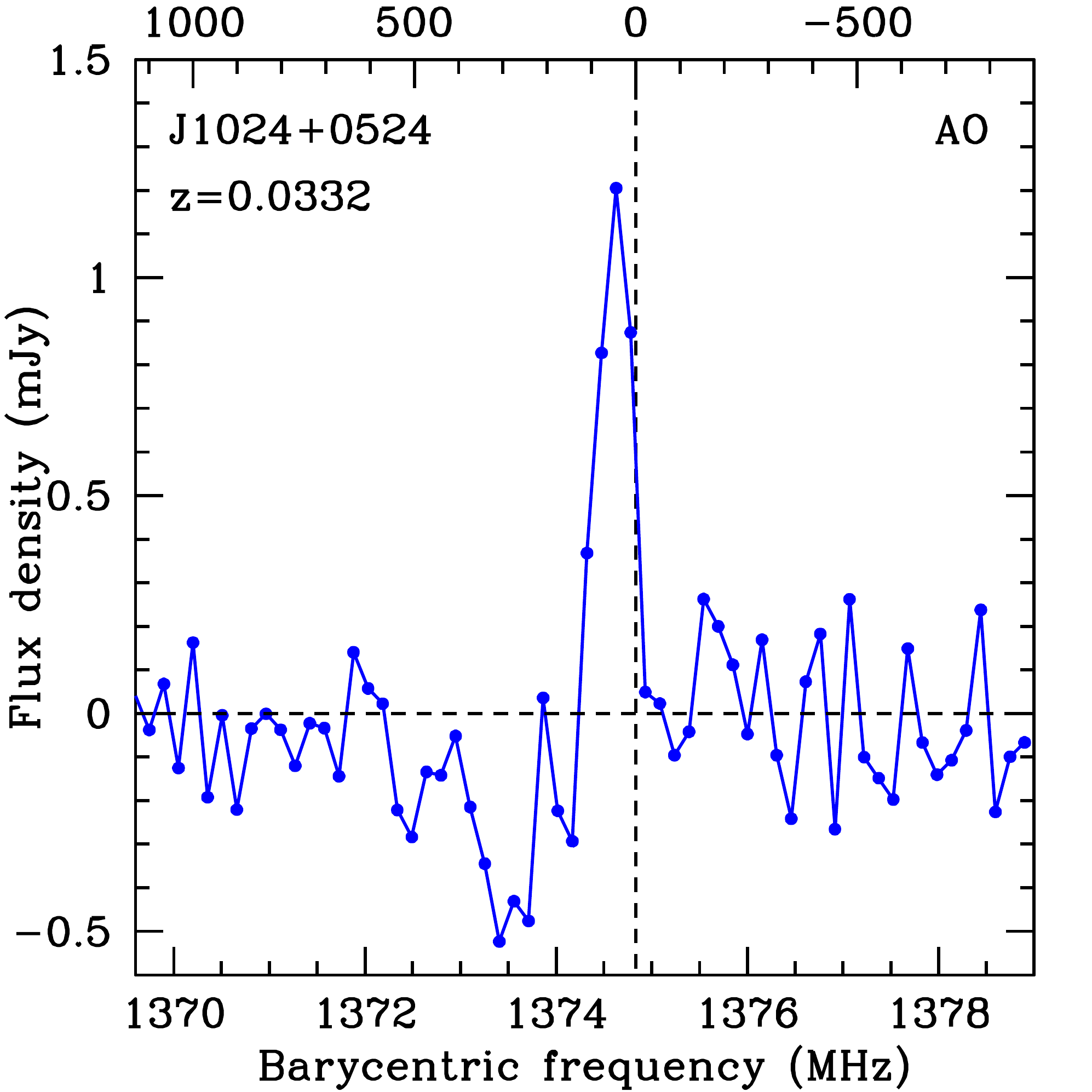}
\includegraphics[width=1.7in]{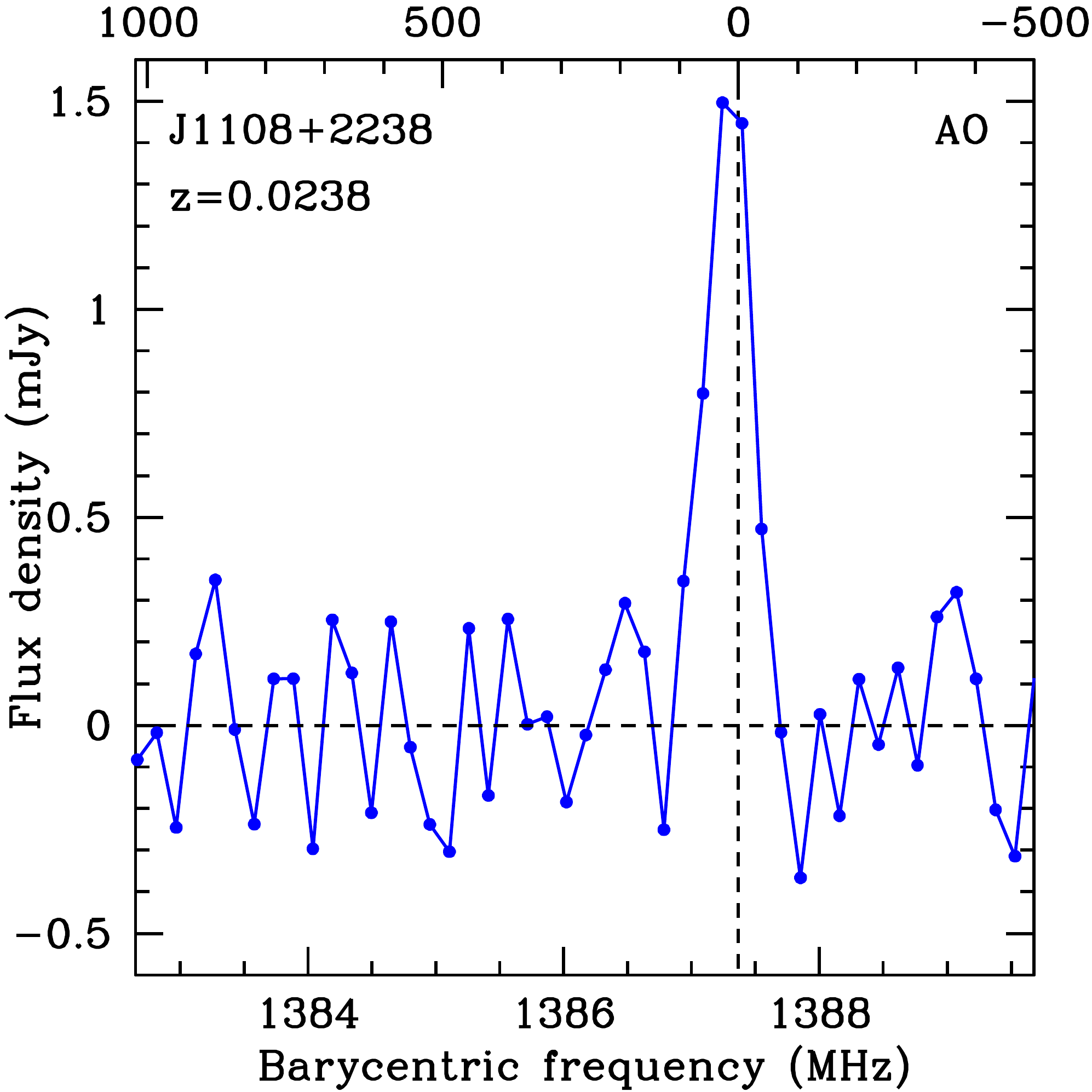}
\includegraphics[width=1.7in]{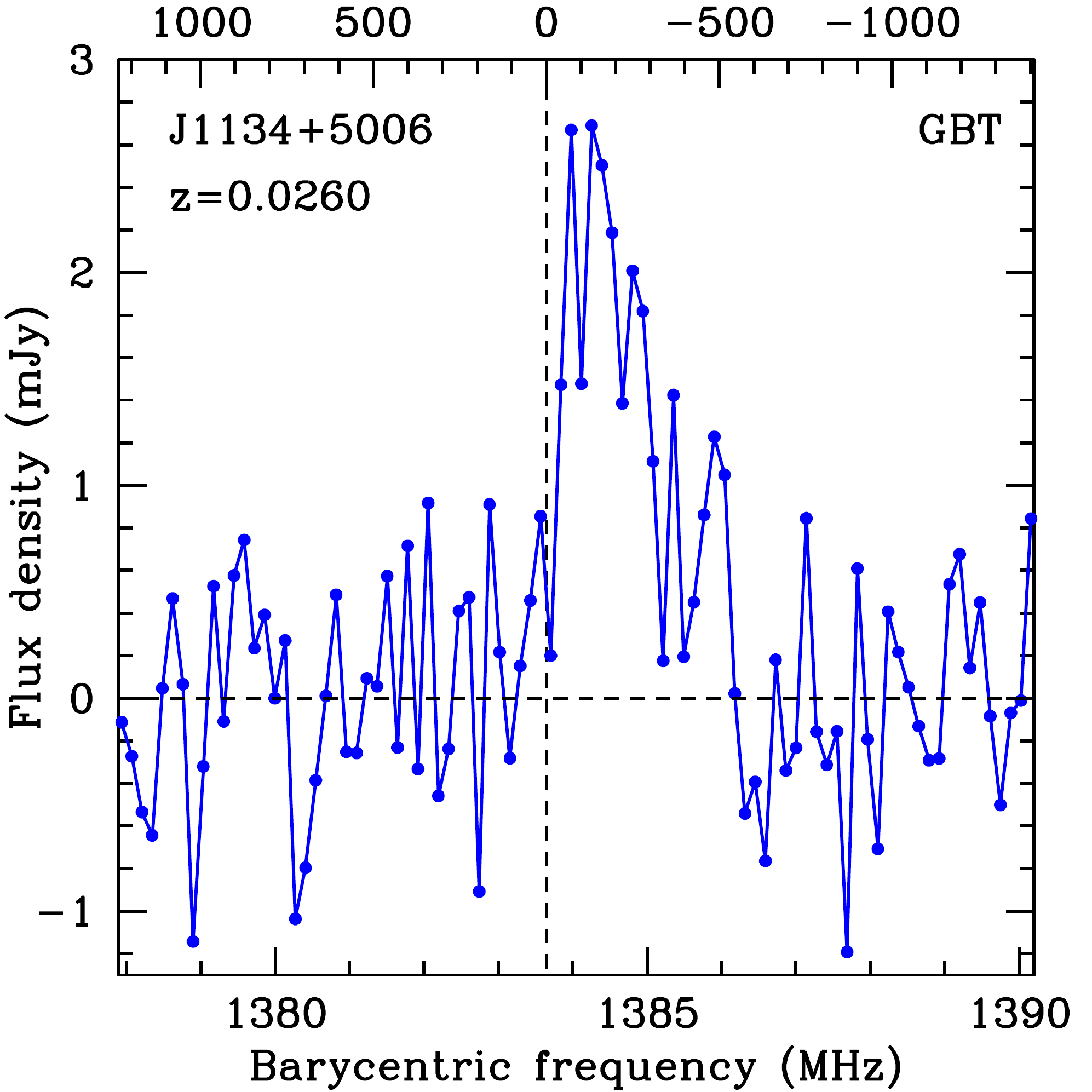}
\includegraphics[width=1.7in]{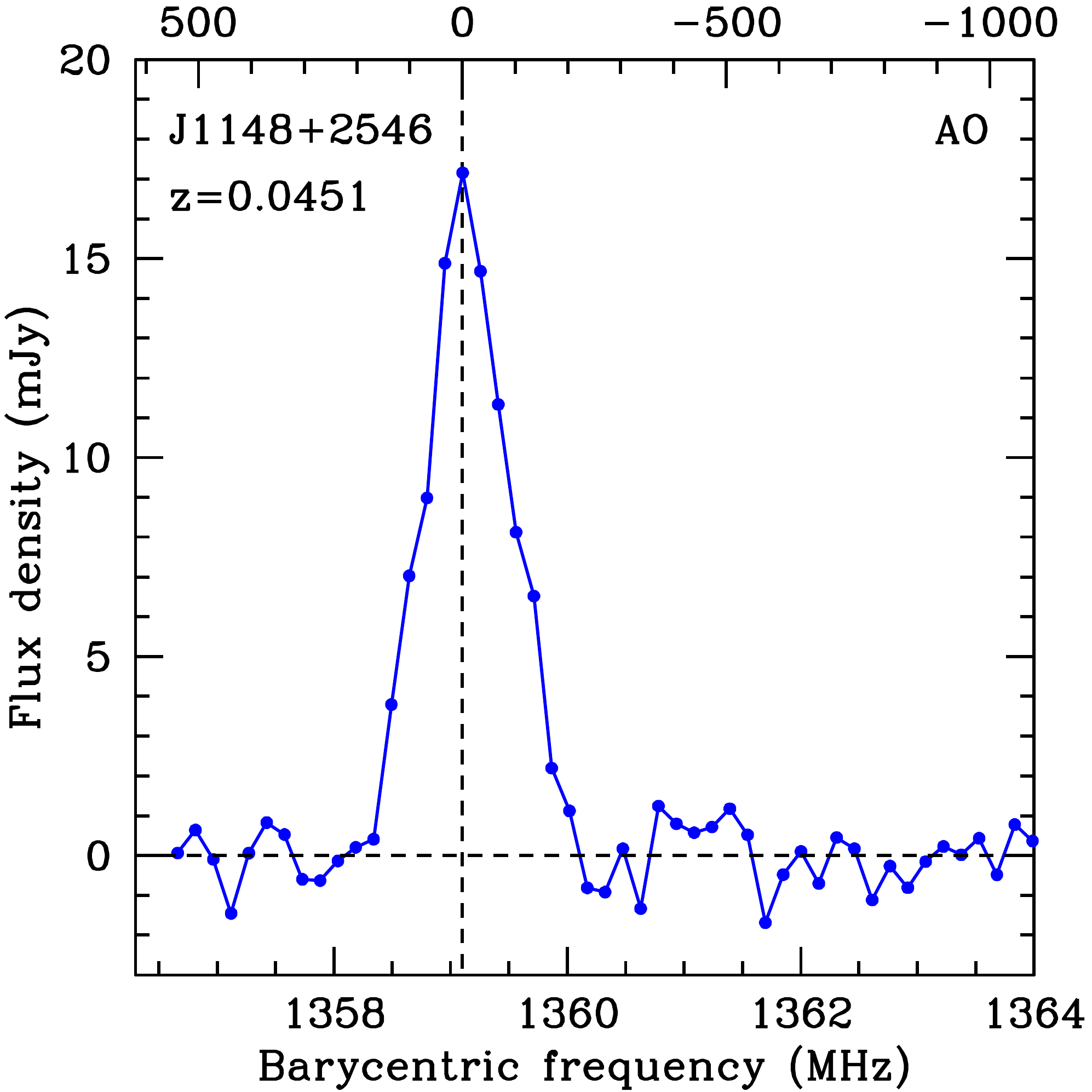}
\includegraphics[width=1.7in]{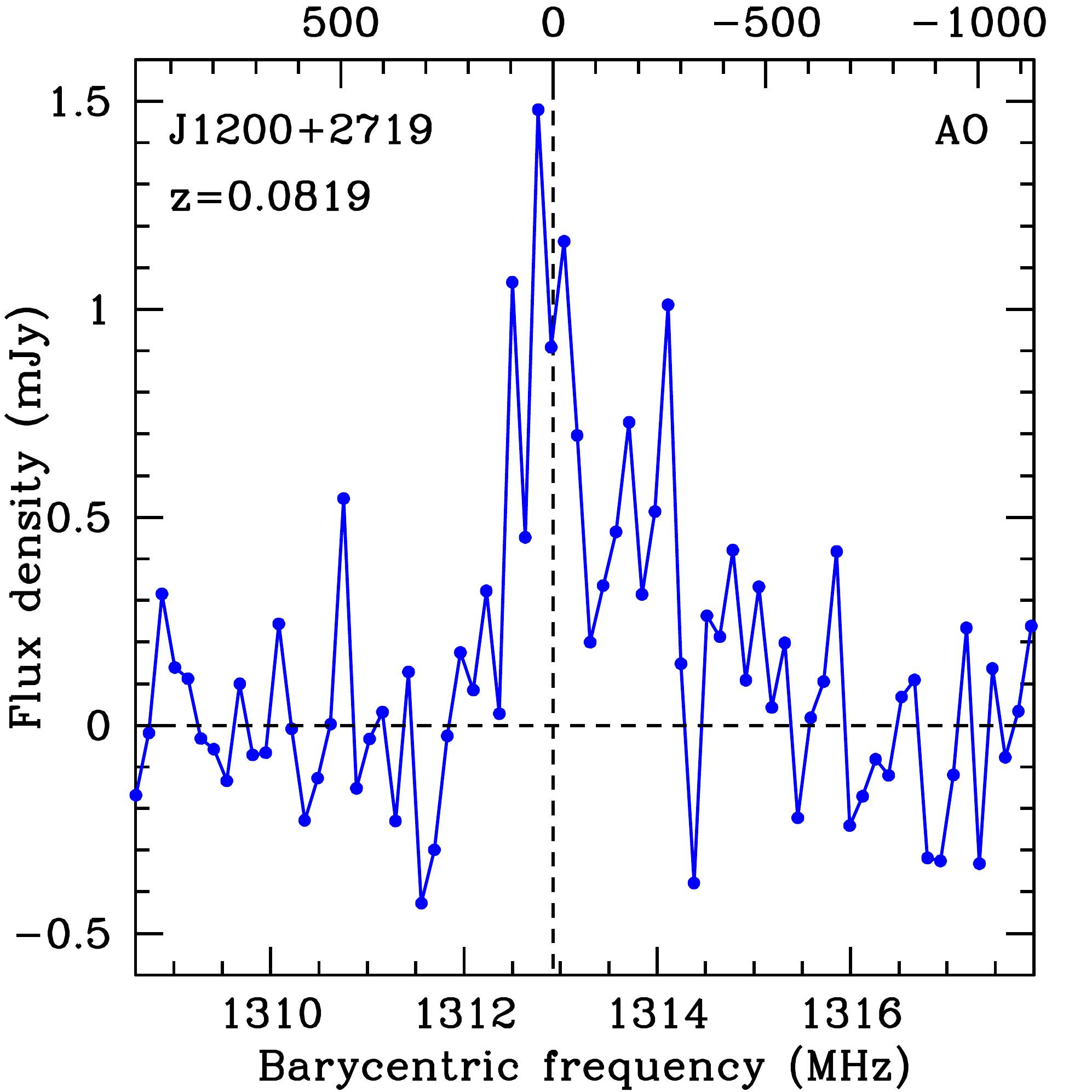}
\includegraphics[width=1.7in]{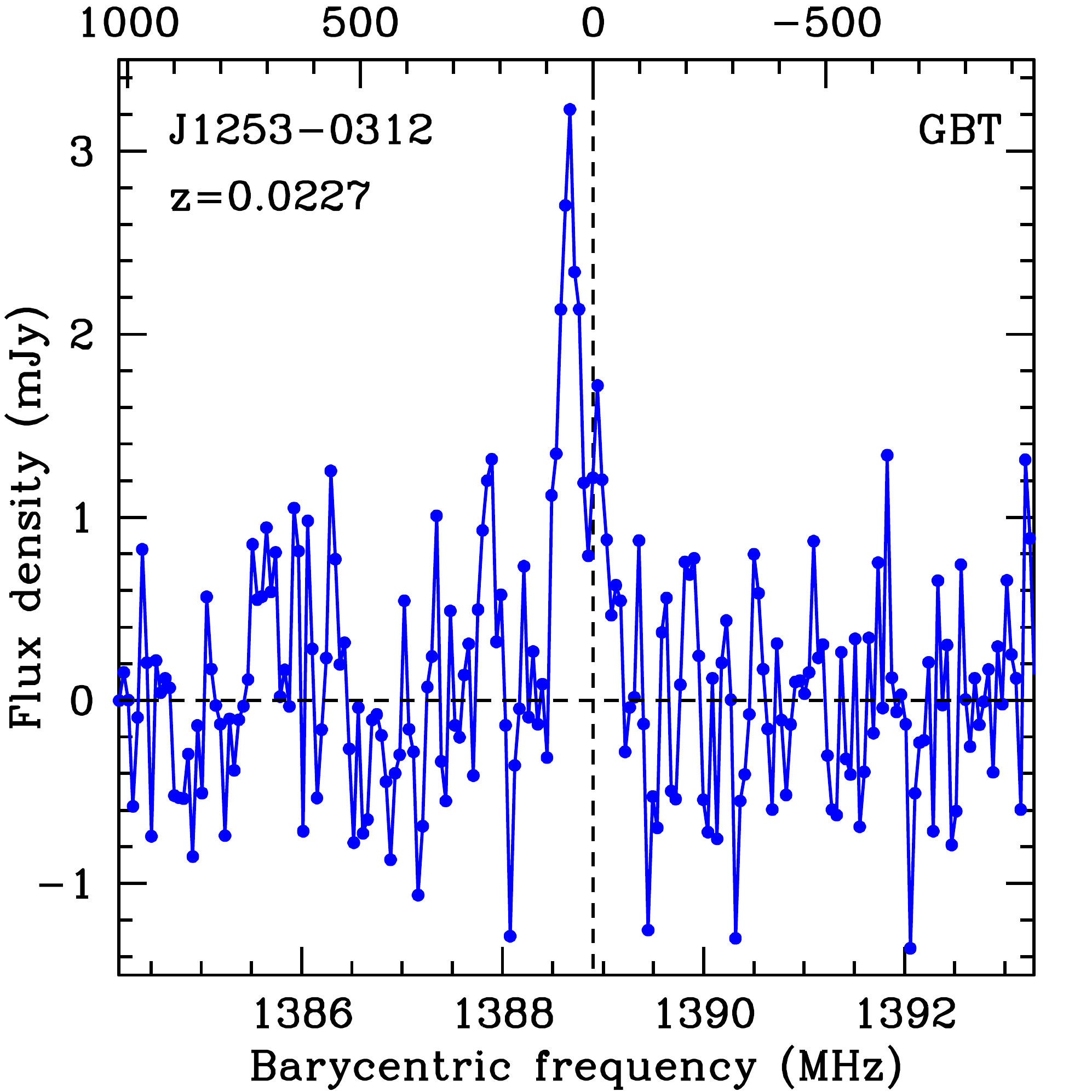}
\includegraphics[width=1.7in]{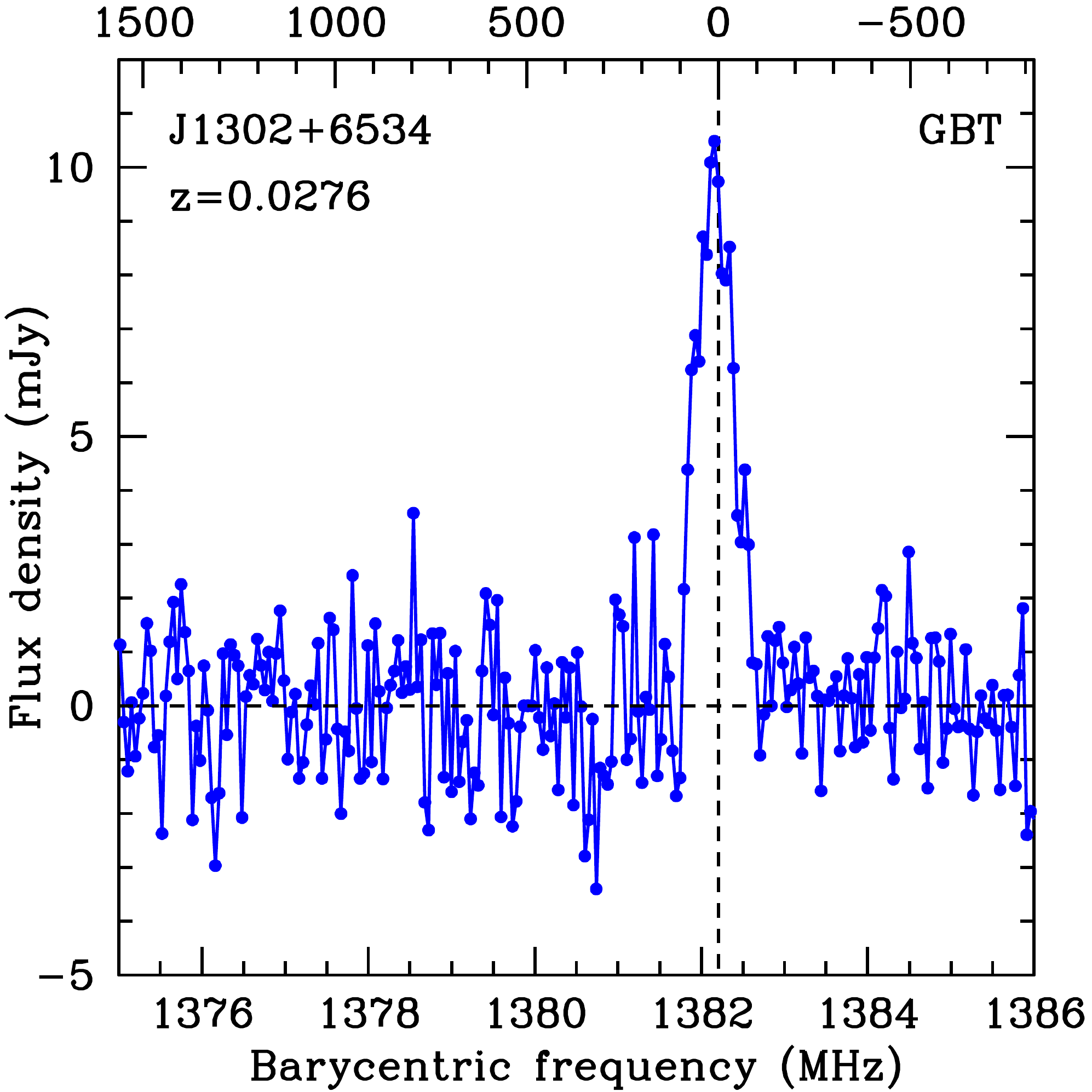}
\includegraphics[width=1.7in]{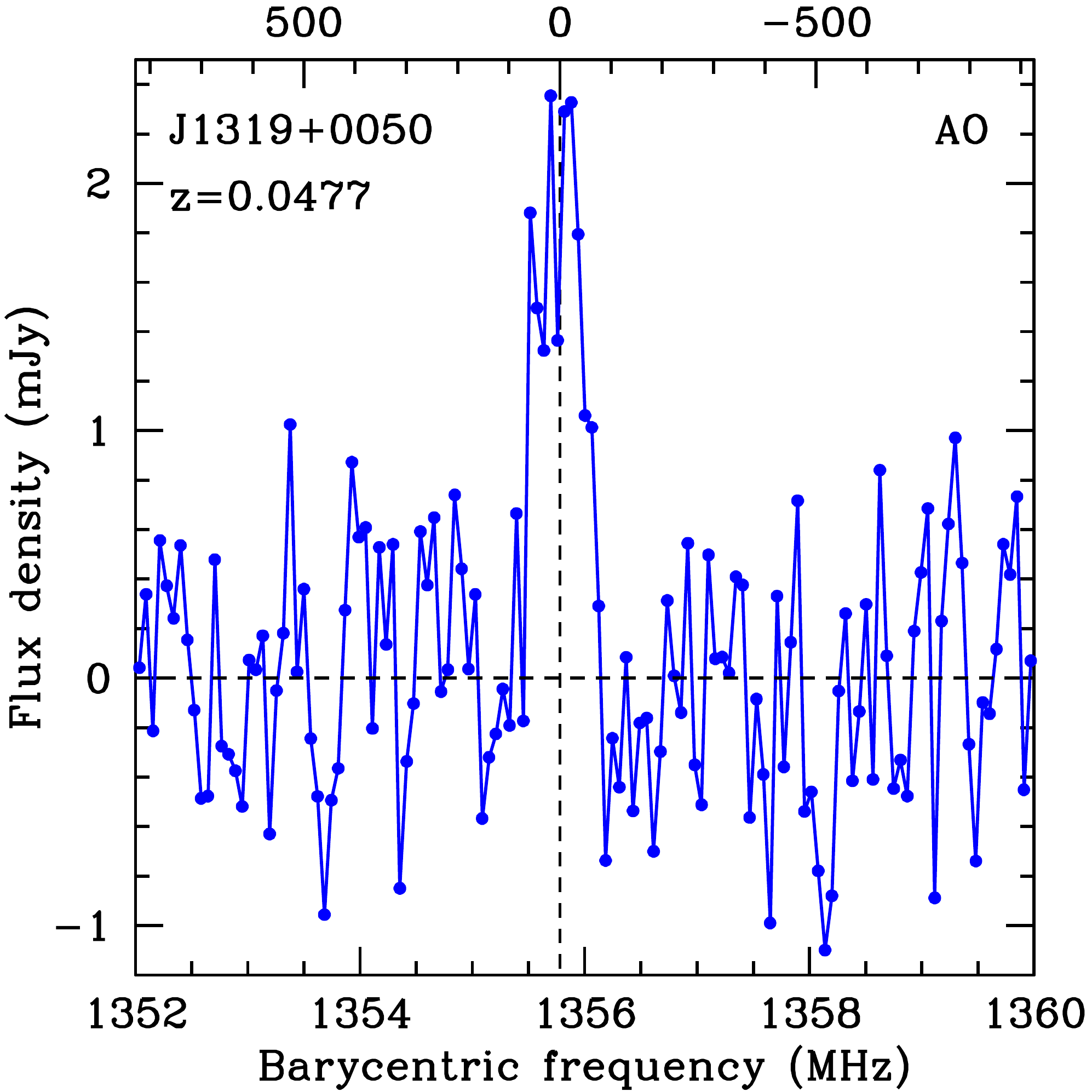}
\includegraphics[width=1.7in]{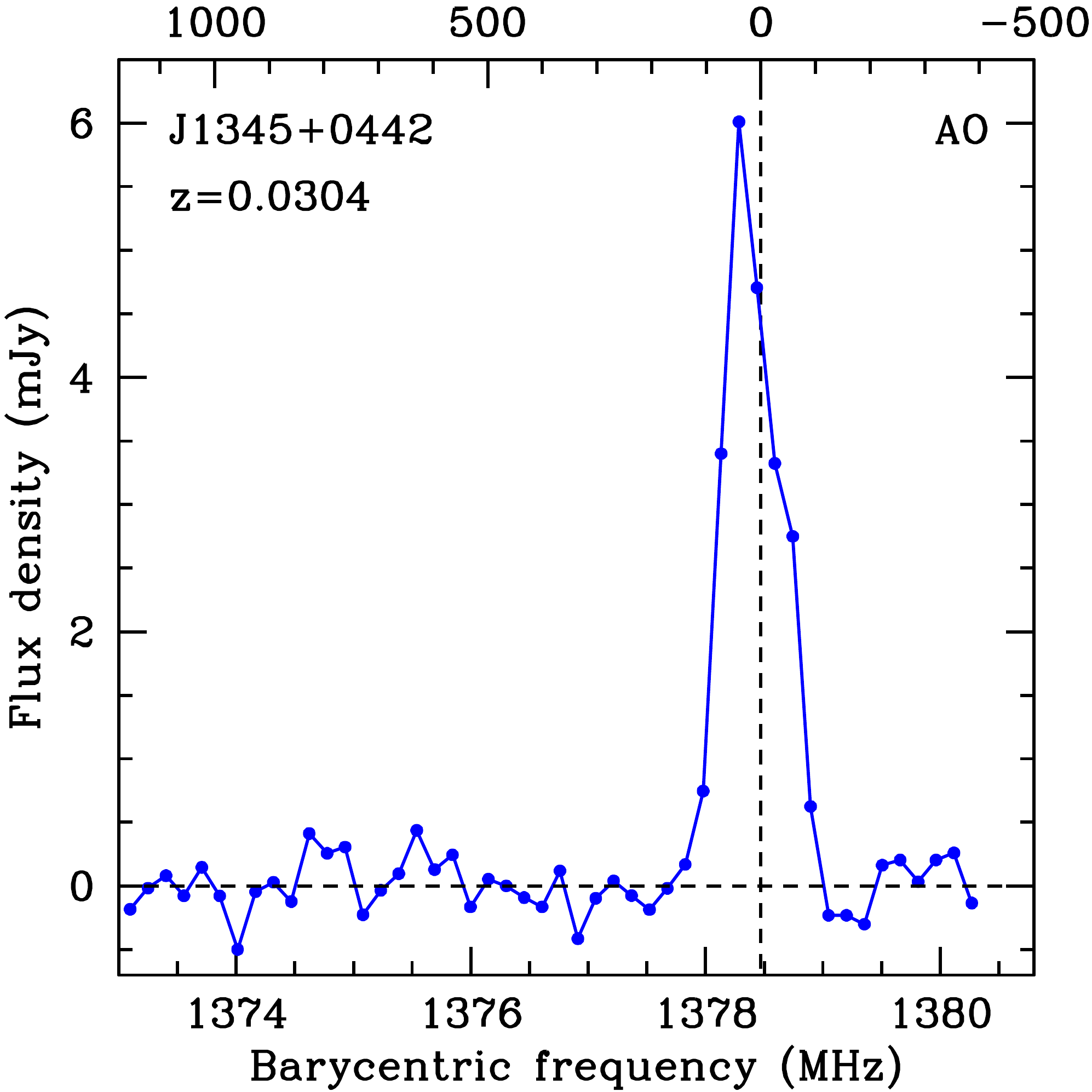}
\includegraphics[width=1.7in]{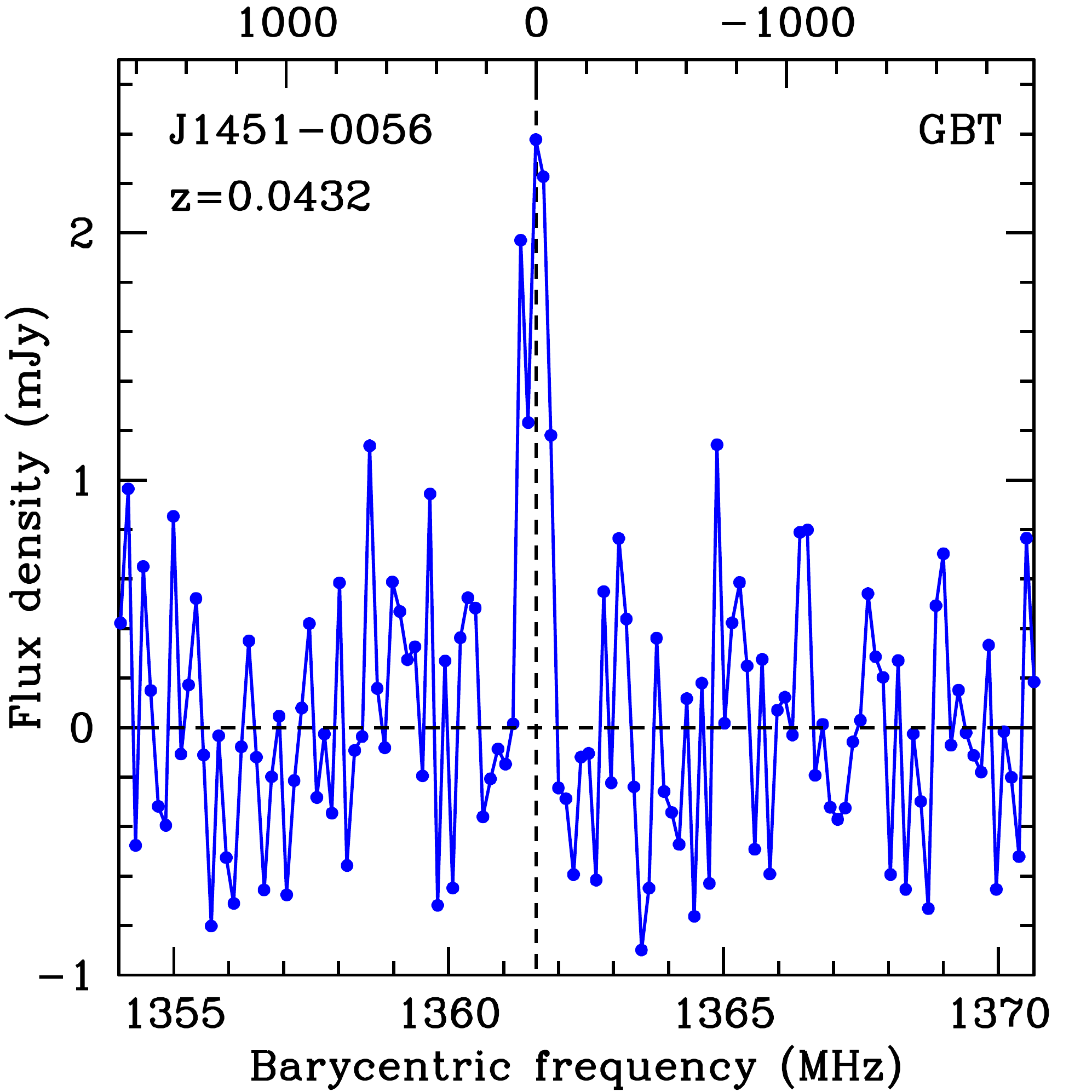}
\includegraphics[width=1.7in]{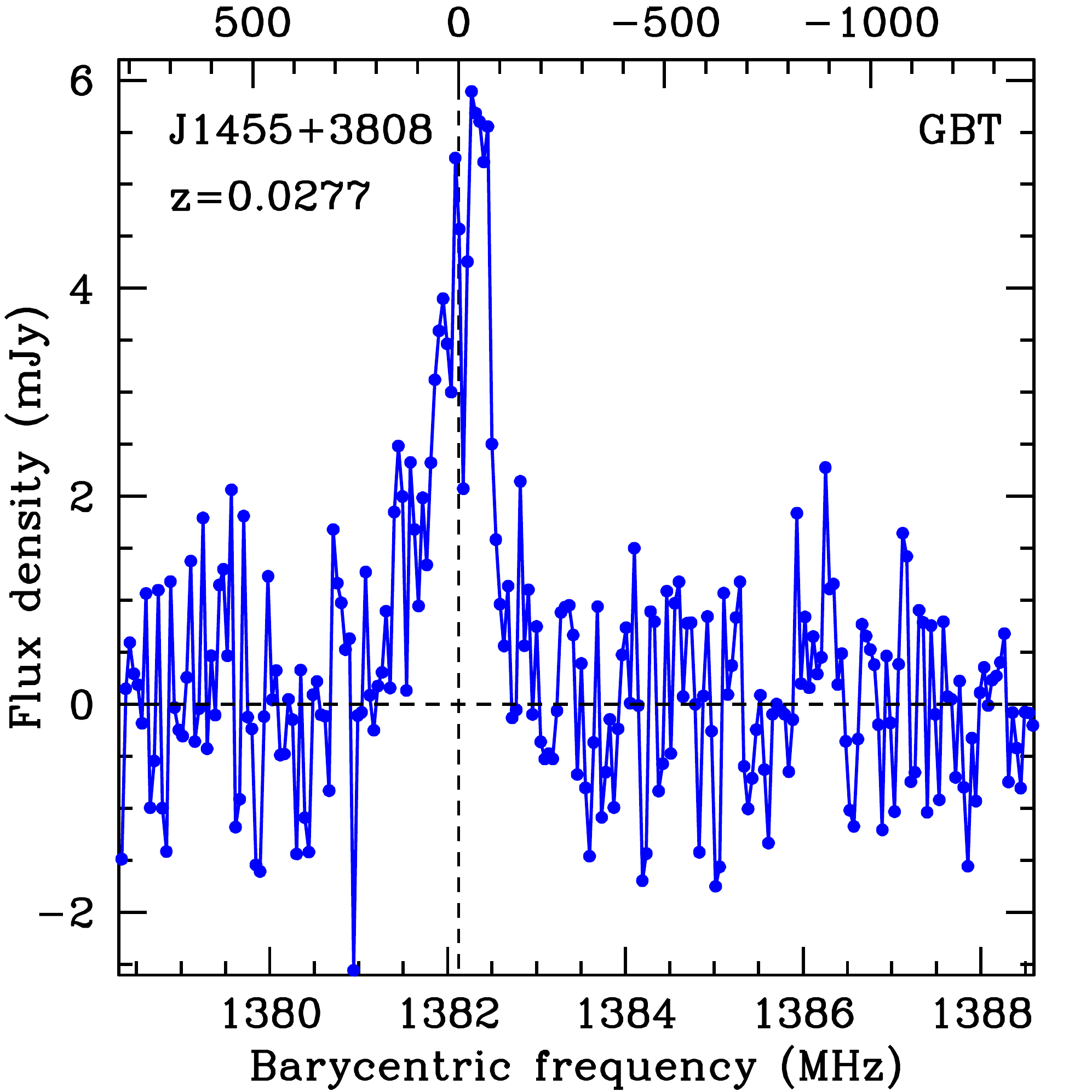}
\includegraphics[width=1.7in]{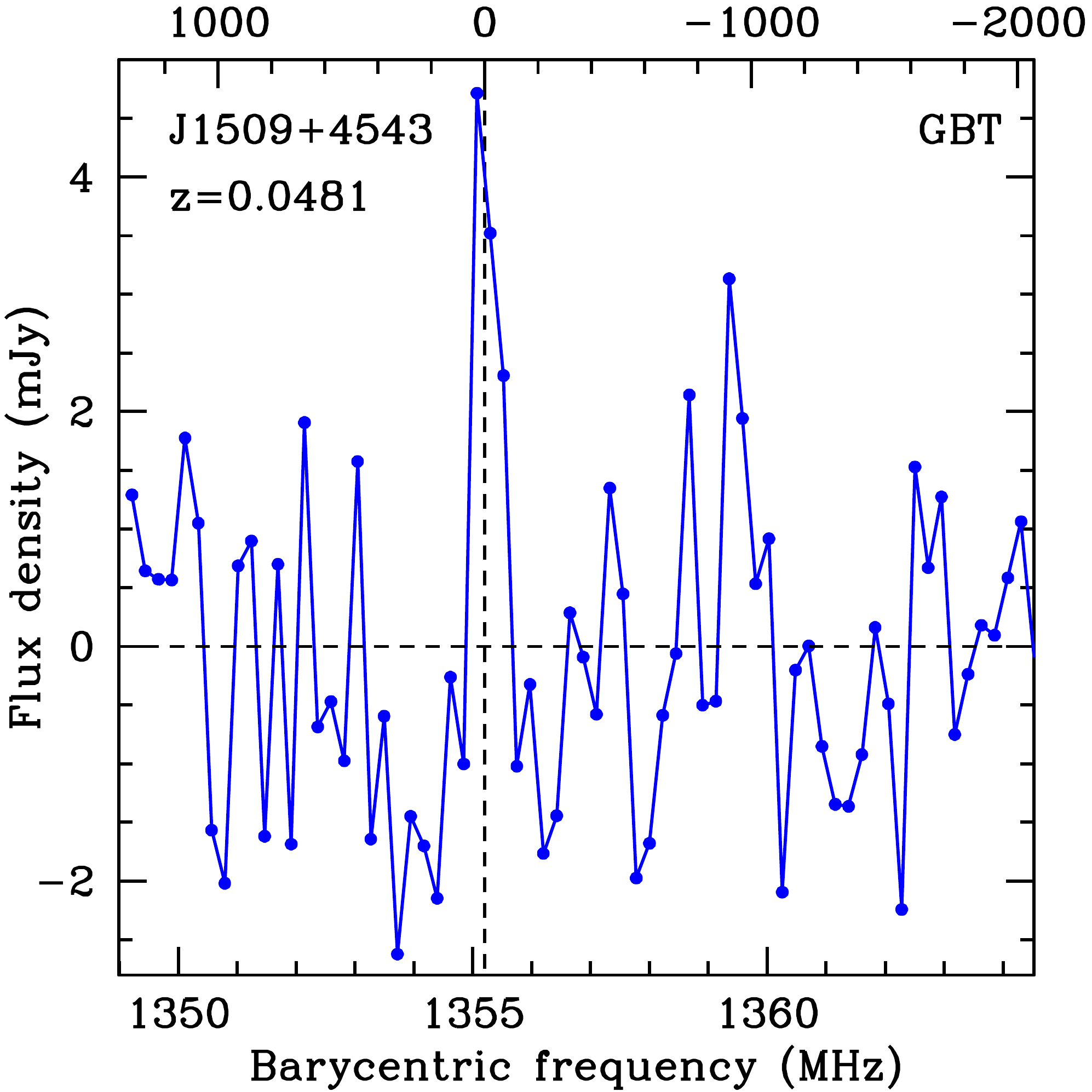}
\includegraphics[width=1.7in]{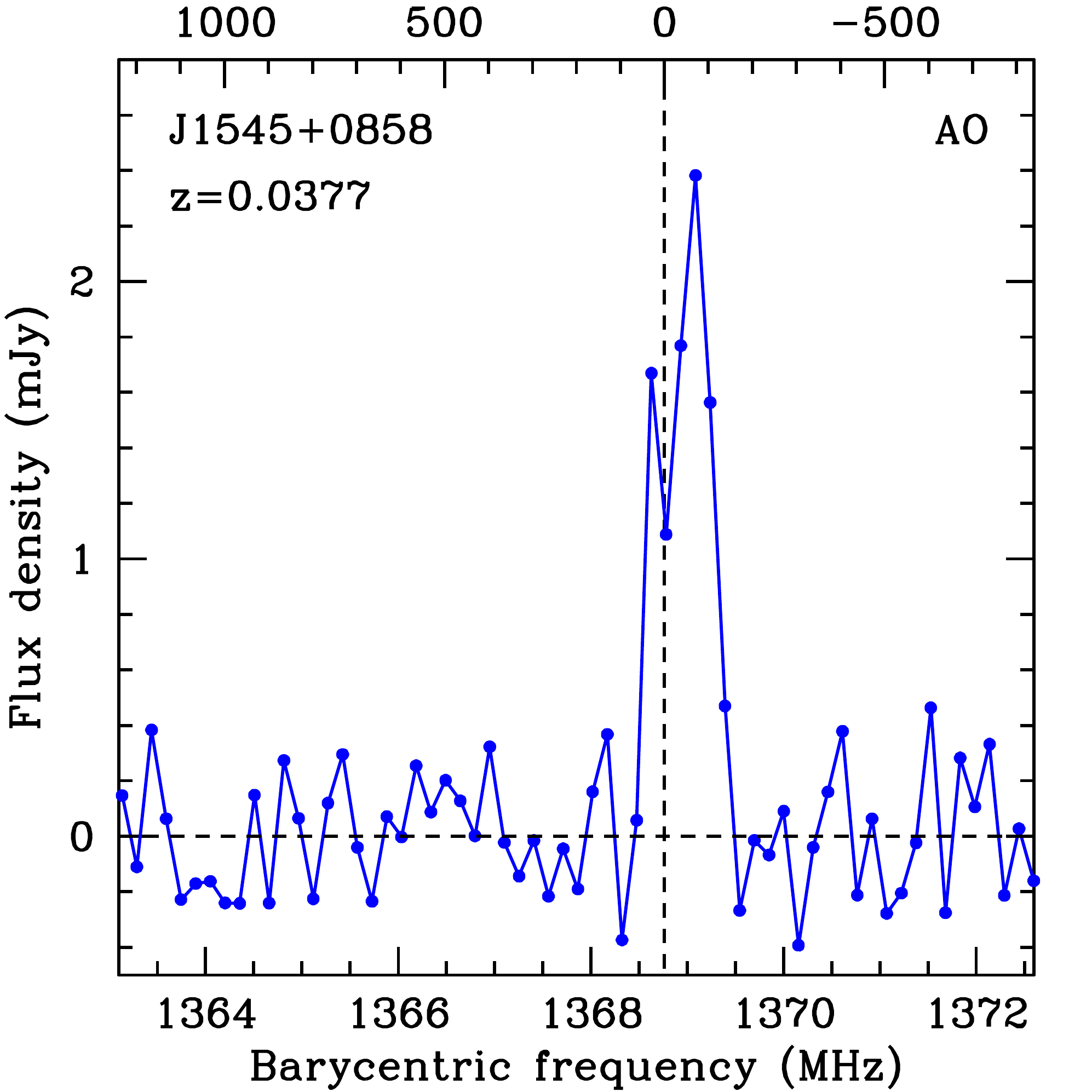}
\includegraphics[width=1.7in]{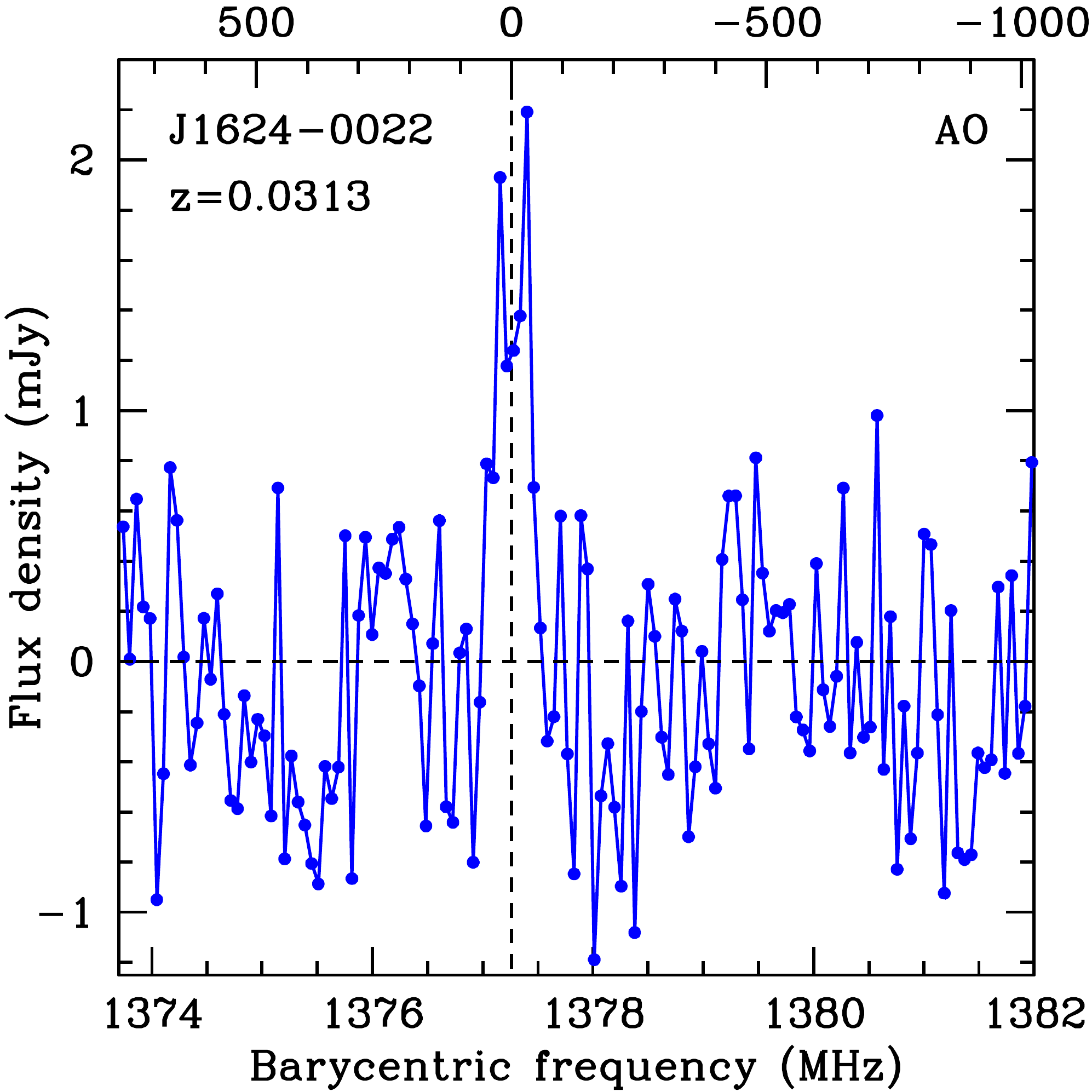}
\includegraphics[width=1.7in]{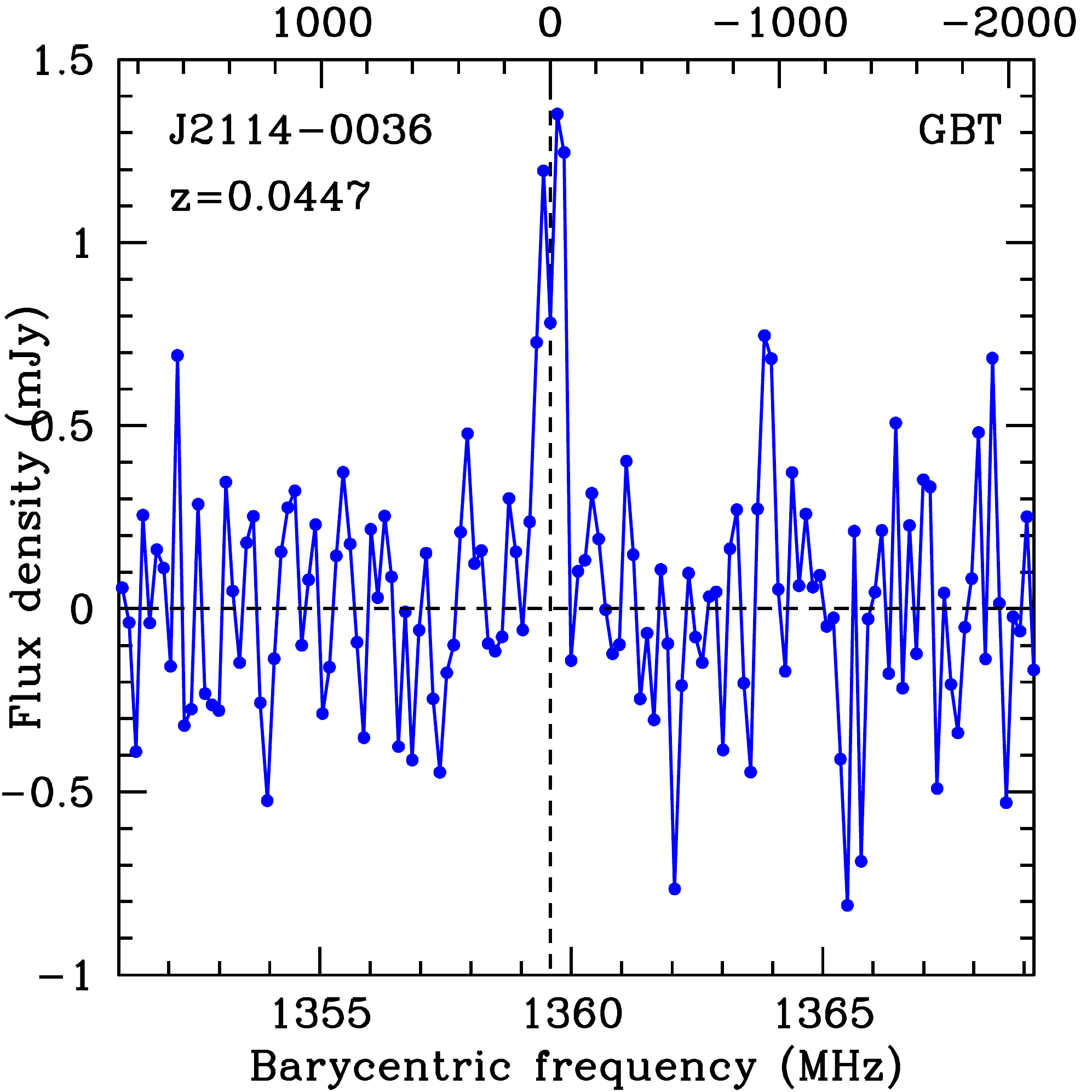}
\caption{The \hii\ emission profiles of the 19 Green Peas with \hii\ detections, ordered 
by Right Ascension. In each panel, the x-axis is barycentric frequency, in MHz; the top
of the panel shows velocity, in \kms, relative to the Green Pea redshift (based on the optical spectra). 
The \hii\ spectra have been smoothed to, and re-sampled at, velocity resolutions of $\approx 10-30$~\kms. Note that the 
\hii\ detections in J0844+0226 and J1010+1255 have $\approx 5\sigma$ significance, and so should
be treated as tentative detections.
\label{fig:spc}}
\end{figure*}

\hii\ emission was detected from 19 Green Peas at $\geq 5\sigma$ significance (two of which, in J0844+0226 and J1010+1255,
have $\approx 5\sigma$ significance and hence should be viewed as tentative detections); the \hii\ spectra of 
these galaxies are shown in Fig.~\ref{fig:spc}. 21 galaxies showed no clear signature of \hii\ emission. 
Table~\ref{table:table1} summarizes the results of the Arecibo and GBT observations; we also include the relevant galaxy properties of each Green Pea, derived from the optical imaging and spectroscopy \citep[e.g.][]{jiang19}. The upper limits 
are computed assuming a Gaussian line profile with a full width at half maximum (FWHM) of 50~\kms, typical
of dwarf galaxies \citep[the \hi\ mass limits are mostly $\lesssim 10^9 \; \msun$, i.e. in the dwarf galaxy range; 
e.g.][]{begum08}. We note that the 
errors quoted on the \hii\ line flux densities and the \hi\ masses are statistical errors, and do not 
include the uncertainty in the flux scale; we estimate this uncertainty to be typically $\approx 10$\%. 

\begin{table*}
\centering
\caption{Results. The columns are (1)~the Green Pea galaxy identifier, (2)~the galaxy redshift, (3)~the telescope used for the observations (Arecibo~$\equiv 1$, GBT~$\equiv 2$, 
(4)~the expected redshifted \hii\ line frequency, (5)~the velocity-integrated \hii\ line flux density (and error), or $3\sigma$ upper limits to this quantity, in Jy~\kms, 
(6)~the inferred \hi\ mass (and error), in units of $10^8 \; \msun$, (7)~the star formation rate (SFR), 
in $\msun$~yr$^{-1}$ \citep{jiang19}, (8)~the stellar mass, in units of $10^8 \; \msun$ \citep{jiang19}, (9)~the \hi-to-stellar mass ratio, $\fhi \equiv \mhi/\mstar$, (10)~the \hi\ 
depletion time, in Gyr, (11)~the absolute blue magnitude, $\mb$, and (12)~the ratio of the luminosities in the [O{\sc iii}]$\lambda$5007 and [O{\sc ii}]$\lambda$3727 lines, 
O32~$\equiv$~[O{\sc iii}$]\lambda$5007/[O{\sc ii}$]\lambda$3727.
\label{table:table1}}
\begin{tabular}{|c|c|c|c|c|c|c|c|c|c|c|c|c|}
\hline
Green Pea & $z$ & Tel. & $\nu_{\rm 21\,cm}$ & $\int S {\rm d}V$ & $\mhi$ & SFR & $\mstar$ & $\fhi$ & $\tau_{\rm dep}$ &  $\mb$ & O32 \\
identifier &    &      &  MHz      & Jy~km~s$^{-1}$ & $ 10^8 \; \msun$  & $\msun$~yr$^{-1}$  & $10^8 \; \msun$ &       &  Gyr             &        &                      \\
\hline
J0007+0226 & $0.0636$ & $1$ & $ 1335.52 $ & $ < 0.062$         & $ < 12 $         & $0.69$ & $   3.0 $ & $< 4.0 $ & $< 1.7$   & $-18.64$ & $48.5$  \\
J0036+0052 & $0.0282$ & $1$ & $ 1381.42 $ & $0.585 \pm 0.034$  & $22.0 \pm 1.3$   & $0.17$ & $  1.7  $ & $12.8$   & $13.0 $   & $-16.41$ & $7.7 $  \\
J0159+0751 & $0.0611$ & $1$ & $ 1338.65 $ & $ < 0.038$         & $ < 6.9$         & $1.0 $ & $  0.49 $ & $< 14.1$ & $< 0.67 $ & $-17.63$ & $60.5$  \\
J0213+0056 & $0.0399$ & $2$ & $ 1365.88 $ & $0.419 \pm 0.022$  & $31.7 \pm 1.7$   & $1.4 $ & $  7.8  $ & $4.1 $   & $2.3  $   & $-17.62$ & $ 8.8$  \\
J0801+3823 & $0.0376$ & $2$ & $ 1368.89 $ & $ < 0.051$         & $ < 3.4$         & $0.57$ & $  15.6 $ & $< 0.22$ & $< 0.60 $ & $-16.08$ & $3.3 $  \\
J0808+1728 & $0.0442$ & $1$ & $ 1360.28 $ & $ < 0.044$         & $ < 4.1$         & $0.45$ & $   2.9 $ & $< 1.4 $ & $< 0.91 $ & $-17.84$ & $14.1$  \\
J0844+0226 & $0.0911$ & $1$ & $ 1301.81 $ & $0.065 \pm 0.013$  & $26.1 \pm 5.4$   & $14.0$ & $ 125.9 $ & $0.21$   & $0.19 $   & $-19.65$ & $4.1 $  \\
J0852+1216 & $0.0759$ & $1$ & $ 1320.19 $ & $ < 0.050$         & $ < 14 $         & $13.4$ & $  73.0 $ & $< 0.19$ & $< 0.10 $ & $-18.32$ & $4.0 $  \\
J0942+4110 & $0.0460$ & $2$ & $ 1357.97 $ & $ < 0.125$         & $ < 13$          & $2.6 $ & $   6.7 $ & $< 1.9 $ & $< 0.49 $ & $-18.92$ & $11.5$  \\
J1010+1255 & $0.0613$ & $1$ & $ 1338.31 $ & $0.053 \pm 0.011$  & $9.6 \pm 2.0$    & $5.3 $ & $  6.7  $ & $1.4 $   & $0.18 $   & $-20.02$ & $4.2 $  \\
J1015+3054 & $0.0918$ & $1$ & $ 1301.04 $ & $ < 0.037$         & $ < 15 $         & $6.1 $ & $  19.2 $ & $< 0.78$ & $< 0.25 $ & $-19.88$ & $2.3 $  \\
J1024+0524 & $0.0332$ & $1$ & $ 1374.76 $ & $0.074 \pm 0.014$  & $3.85 \pm 0.73$  & $1.6 $ & $  0.95 $ & $4.0 $   & $0.25 $   & $-18.91$ & $5.6 $  \\
J1108+2238 & $0.0238$ & $1$ & $ 1387.37 $ & $0.155 \pm 0.016$  & $4.14 \pm 0.43$  & $0.58$ & $  5.6  $ & $0.74$   & $0.71 $   & $-16.89$ & $2.8 $  \\
J1134+5006 & $0.0260$ & $2$ & $ 1384.44 $ & $0.799 \pm 0.065$  & $25.4 \pm 2.1$   & $2.2 $ & $  0.98 $ & $25.8$   & $1.1  $   & $-18.24$ & $2.5 $  \\
J1148+2546 & $0.0451$ & $1$ & $ 1359.07 $ & $3.182 \pm 0.078$  & $309.0 \pm 7.6$  & $5.2 $ & $  5.7  $ & $54.7$   & $6.0  $   & $-19.52$ & $5.4 $  \\
J1200+2719 & $0.0819$ & $1$ & $ 1312.91 $ & $0.310 \pm 0.028$  & $100.6 \pm 9.0$  & $3.8 $ & $ 20.0  $ & $4.6 $   & $2.5  $   & $-18.83$ & $12.9$  \\
J1224+0105 & $0.0398$ & $2$ & $ 1365.99 $ & $ < 0.063$         & $ < 4.8$         & $0.85$ & $  15.0 $ & $< 0.32$ & $< 0.56 $ & $-17.12$ & $3.4 $  \\
J1224+3724 & $0.0404$ & $2$ & $ 1365.25 $ & $ < 0.077$         & $ < 5.9$         & $0.96$ & $   4.0 $ & $< 1.5 $ & $< 0.62 $ & $-17.87$ & $8.6 $  \\
J1226+0415 & $0.0942$ & $1$ & $ 1298.10 $ & $ < 0.073$         & $ < 32 $         & $5.1 $ & $  29.0 $ & $< 1.1 $ & $< 0.62 $ & $-19.99$ & $11.2$  \\
J1253-0312 & $0.0227$ & $2$ & $ 1388.89 $ & $0.235 \pm 0.021$  & $56.9 \pm 5.1$   & $89.2$ & $  1.3  $ & $41.8$   & $0.062$   & $-19.58$ & $4.6 $  \\
J1302+6534 & $0.0276$ & $2$ & $ 1382.20 $ & $1.179 \pm 0.052$  & $42.6 \pm 1.9$   & $0.69$ & $ 11.1  $ & $3.8 $   & $6.2  $   & $-17.51$ & $3.9 $  \\
J1319+0050 & $0.0477$ & $1$ & $ 1355.78 $ & $0.234 \pm 0.025$  & $21.7 \pm 2.3$   & $1.4 $ & $  9.8  $ & $2.2 $   & $1.6  $   & $-17.99$ & $2.7 $  \\
J1329+1700 & $0.0942$ & $1$ & $ 1298.16 $ & $ < 0.086$         & $ < 37 $         & $9.9 $ & $  49.1 $ & $< 0.75$ & $< 0.37 $ & $-18.65$ & $4.0 $  \\
J1345+0442 & $0.0304$ & $1$ & $ 1378.47 $ & $0.650 \pm 0.026$  & $28.5 \pm 1.1$   & $1.1 $ & $  7.2  $ & $4.0 $   & $2.6  $   & $-17.80$ & $3.0 $  \\
J1359+5726 & $0.0338$ & $2$ & $ 1373.93 $ & $ < 0.095$         & $ < 5.1$         & $2.1 $ & $  13.3 $ & $< 0.38$ & $< 0.24 $ & $-17.28$ & $3.6 $  \\
J1411+0556 & $0.0493$ & $2$ & $ 1353.62 $ & $ < 0.047$         & $ < 5.5$         & $1.4 $ & $   1.4 $ & $< 3.9 $ & $< 0.40 $ & $-19.88$ & $19.6$  \\
J1423+2257 & $0.0328$ & $1$ & $ 1375.24 $ & $ < 0.025$         & $ < 1.3$         & $0.98$ & $   1.5 $ & $< 0.89$ & $< 0.13 $ & $-17.16$ & $8.5 $  \\
J1432+5152 & $0.0256$ & $2$ & $ 1384.94 $ & $ < 0.079$         & $ < 2.4$         & $0.58$ & $  9.5  $ & $< 0.26$ & $< 0.43 $ & $-17.38$ & $3.3 $  \\
J1448-0110 & $0.0274$ & $2$ & $ 1382.50 $ & $ < 0.054$         & $ < 1.9$         & $2.8 $ & $  0.84 $ & $< 2.3 $ & $< 0.068$ & $-18.62$ & $10.2$  \\
J1451-0056 & $0.0432$ & $2$ & $ 1361.59 $ & $0.288 \pm 0.029$  & $25.6 \pm 2.6$   & $0.63$ & $  11.1 $ & $2.3 $   & $4.0  $   & $-18.38$ & $3.5 $  \\
J1455+3808 & $0.0277$ & $2$ & $ 1382.13 $ & $0.855 \pm 0.047$  & $31.0 \pm 1.7$   & $0.96$ & $   2.2 $ & $14.4$   & $3.2  $   & $-17.56$ & $7.5 $  \\
J1509+3731 & $0.0326$ & $2$ & $ 1375.58 $ & $< 0.088$          & $< 4.4$          & $1.77$ & $  0.86 $ & $< 5.2$  & $< 0.25$  & $-18.41$ & $19.2$  \\
J1509+4543 & $0.0481$ & $2$ & $ 1355.18 $ & $0.543 \pm 0.081$  & $60.0 \pm 8.9$   & $3.3 $ & $  37.8 $ & $1.6 $   & $1.8  $   & $-18.66$ & $3.1 $  \\
J1518+1955 & $0.0751$ & $1$ & $ 1321.19 $ & $ < 0.041$         & $ < 11 $         & $4.93$ & $  31.7 $ & $< 0.35$ & $< 0.22 $ & $-19.47$ & $3.3 $  \\
J1545+0858 & $0.0377$ & $1$ & $ 1368.76 $ & $0.302 \pm 0.019$  & $17.8 \pm 1.1$   & $4.4 $ & $  10.1 $ & $1.8 $   & $0.40 $   & $-18.91$ & $9.7 $  \\
J1547+2203 & $0.0314$ & $1$ & $ 1377.15 $ & $ < 0.053$         & $ < 2.5$         & $0.68$ & $   8.5 $ & $< 0.29$ & $< 0.37 $ & $-17.31$ & $5.9 $  \\
J1608+3528 & $0.0327$ & $1$ & $ 1375.38 $ & $ < 0.12$          & $< 0.61$         & $0.46$ & $  0.29 $ & $< 2.1 $ & $< 0.13 $ & $-17.01$ & $51.1$  \\
J1624-0022 & $0.0313$ & $1$ & $ 1377.27 $ & $0.134 \pm 0.021$  & $62.2 \pm 9.8$   & $4.0 $ & $  13.0 $ & $4.8 $   & $1.6  $   & $-17.22$ & $5.0 $  \\
J2114-0036 & $0.0447$ & $2$ & $ 1359.59 $ & $0.173 \pm 0.020$  & $16.5 \pm 1.9$   & $0.79$ & $   2.2 $ & $7.6 $   & $2.1  $   & $-19.67$ & $9.3 $  \\
J2302+0049 & $0.0331$ & $1$ & $ 1374.91 $ & $ < 0.050$         & $ < 2.6$         & $0.49$ & $   1.0 $ & $< 2.5 $ & $< 0.53 $ & $-16.91$ & $11.6$  \\
\hline
\end{tabular}
\vskip 0.1in
\end{table*}

\section{Discussion}
\label{sec:discuss} 

Our Arecibo and GBT \hii\ spectroscopy of Green Pea galaxies has yielded an $\approx 50$\% 
detection rate, with 19 detections of \hii\ emission at redshifts $z \approx 0.023 - 0.091$.
These are the first measurements of the atomic gas content of Green Pea galaxies. The \hi\ masses 
of the detected galaxies lie in the range $\approx (4 - 300) \times 10^8 \; \rm M_\odot$, with a 
median value of $2.6 \times 10^9 \; \rm M_\odot$. For the non-detections, the $3\sigma$ upper limits 
on the \hi\ mass lie in the range $(0.6 - 32) \times 10^8 \; \rm M_\odot$, with a median upper limit 
of $5.5 \times 10^8 \; \rm M_\odot$. Note that the large primary beams of Arecibo
and the GBT imply that we cannot rule out the possibility that some of the \hii\ emission in
the detections may arise from companion galaxies.

\begin{figure*}[t!]
\centering
\includegraphics[width=3.3in]{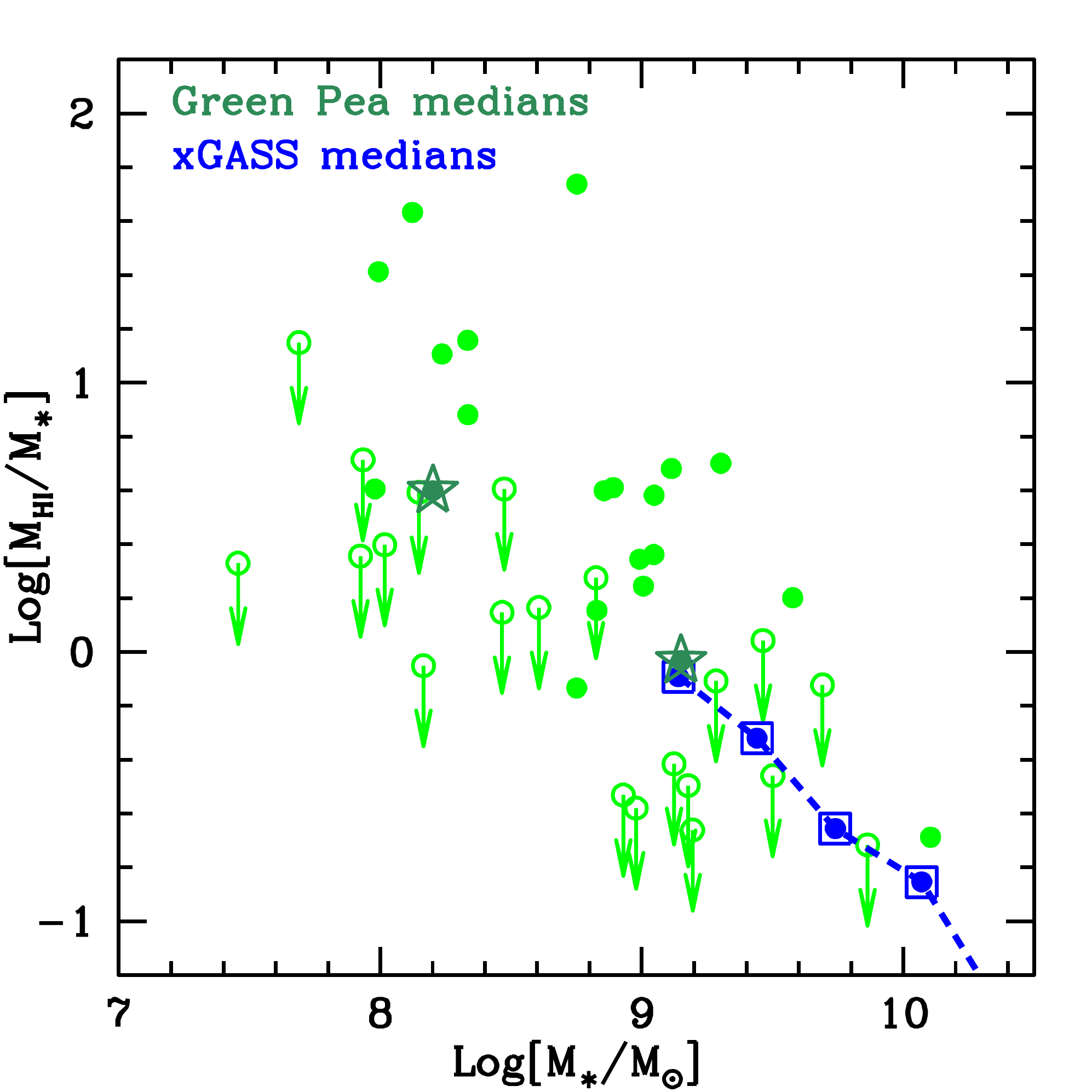}
\includegraphics[width=3.3in]{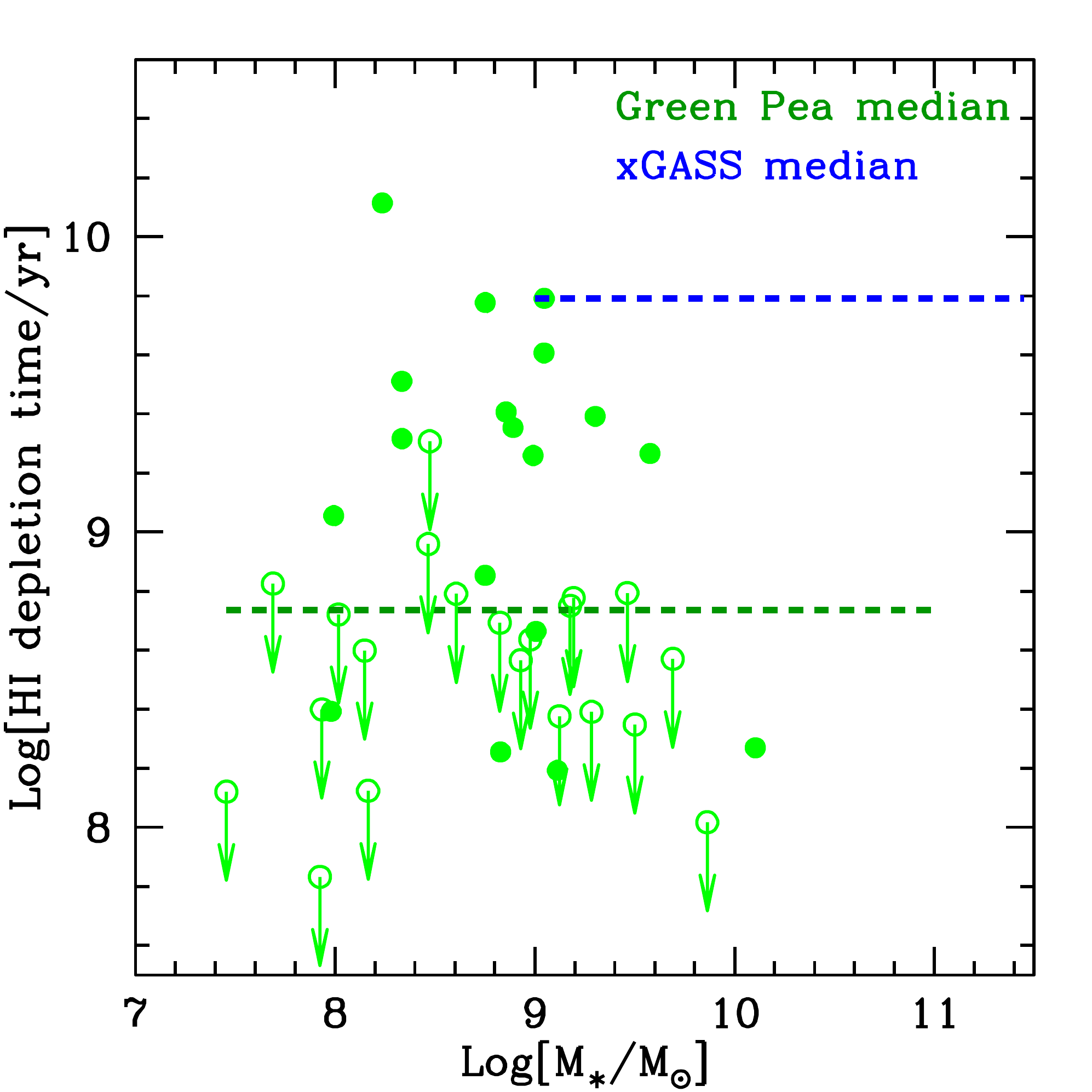}
\caption{[A]~The \hi-to-stellar mass ratio $\fhi \equiv \mhi/\mstar$ plotted against
the stellar mass $\mstar$, for the 40 Green Peas. Detections of \hii\ emission are shown 
as filled (green) circles, and non-detections as open circles with downward-pointing arrows.
The two dark green stars show the median values of $\fhi$ in two stellar mass bins. The 
filled blue circles indicate the median values of $\fhi$ in the xGASS sample \citep{catinella18}.
[B]~The \hi\ depletion time, $\tau_{\rm dep}$, plotted against the stellar mass $\mstar$ for 
the Green Pea galaxies. The dashed lines indicate the median \hi\ depletion timescales for the 
Green Peas (green) and galaxies from the xGASS sample \citep[blue;][]{saintonge17}. The median \hi\ depletion timescale of the Green Peas is seen to be an order of magnitude lower than that of the xGASS galaxies.
\label{fig:hi}}
\end{figure*}

Fig.~\ref{fig:hi}[A] plots the \hi-to-stellar mass ratio $\fhi \equiv \mhi/\mstar$ against
the stellar mass $\mstar$ of the 40 Green Peas of our sample. We used the xGASS sample 
as the comparison  sample, as this is a stellar mass-selected ($\mstar \geq 10^9 \, \msun$)
sample of nearby galaxies, with \hii\ emission studies \citep{catinella18}. The dark green stars 
indicate the median values of $\fhi$ (treating the $3\sigma$ upper limits to $\fhi$ 
as detections) in two stellar mass bins, while the filled blue circles indicate 
the median value of $\fhi$ in galaxies in different stellar mass bins in the xGASS sample 
\citep{catinella18}, with the dashed blue line connecting the xGASS values.
It is clear that the median value of $\fhi$ for Green Peas in the higher $\mstar$ bin is 
in excellent agreement with the median value for xGASS galaxies at the same $\mstar$, 
while the median $\fhi$ in the lower $\mstar$ bin appears to lie close to the extrapolated xGASS 
relation \citep{catinella18}. It thus appears that the \hi\ content of Green Pea galaxies,
relative to their stellar mass, is in excellent agreement with that of ``normal'' 
galaxies in the nearby Universe.

The atomic gas depletion timescale $\tau_{\rm dep} \equiv \mhi/{\rm SFR}$ gives the 
timescale for which a galaxy can continue to form stars without replenishment of its 
\hi\ reservoir. Lower values of $\tau_{\rm dep}$ would imply that a galaxy's star-formation
activity would be regulated by the availability of \hi; for example, \citet{chowdhury20} 
argue that the cause of the decline of the star-formation activity in the Universe 
at $z < 1$ is because the \hi\ reservoirs in star-forming galaxies are not sufficient
to support their star-formation activity for more than $\approx 1-2$~Gyr. However, at 
$z \lesssim 0.35$, the \hi\ depletion timescale has been found to be relatively long in 
main-sequence galaxies, $\approx 5-10$~Gyr \citep[e.g.][]{saintonge17,bera19}. Fig.~\ref{fig:hi}[B] 
plots the $\tau_{\rm dep}$ values 
of our Green Pea galaxies against stellar mass; the dashed green line shows the median
value of the sample, $\tau_{\rm dep, med} \approx 0.58$~Gyr (conservatively treating the upper 
limits on $\mhi$ as detections). For comparison, the median value of $\tau_{\rm dep}$ in the xGASS
sample (again treating upper limits to $\mhi$ as detections), shown by the dashed blue 
line in the figure, is $\approx 6$~Gyr \citep{saintonge17,catinella18}, larger by an order of 
magnitude. 
It appears that the starburst activity in the Green Peas will exhaust their atomic fuel on very short timescales, far 
shorter than in most other galaxies in the nearby Universe.

The depletion time of star-forming material could be longer than the \hi\ depletion timescale when the H$_2$ depletion timescale is taken into account. However, in 
star-forming galaxies at all redshifts, the H$_2$ depletion timescale is typically $\lesssim 1$~Gyr
\citep[e.g.][]{saintonge17,tacconi20}, far shorter than the \hi\ depletion timescale. As such, 
star-formation in such galaxies is not limited by the depletion of \hi, as there is a long timescale
on which the \hi\ can be replenished from the circumgalactic medium. However, the very short 
\hi\ depletion time in Green Peas implies that \hi\ depletion could itself act as a bottleneck 
for star-formation \citep[as has been seen in main-sequence galaxies at $z \approx 1$; ][]{chowdhury20}.

\begin{figure}[t!]
\centering
\includegraphics[width=3.3in]{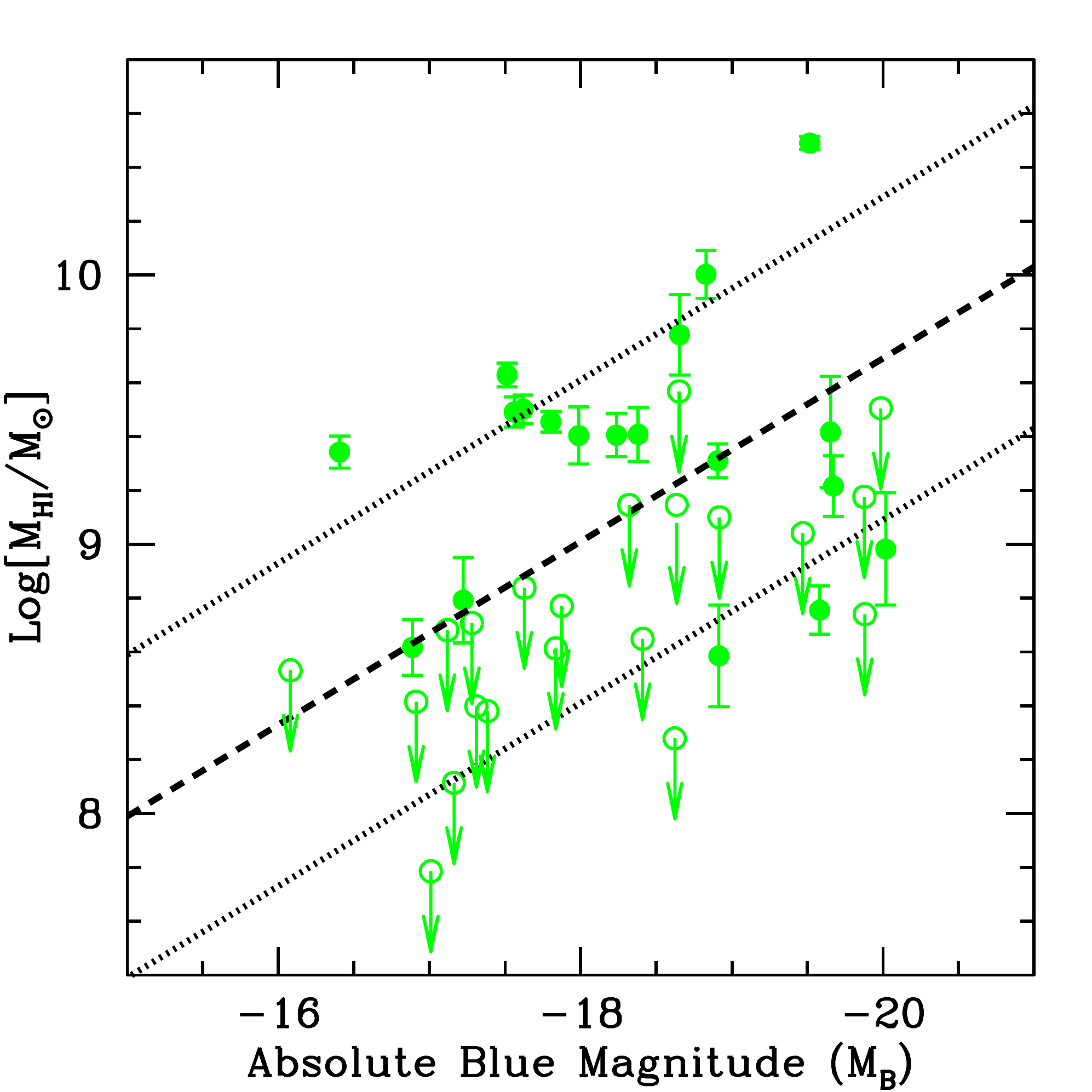}
\includegraphics[width=3.3in]{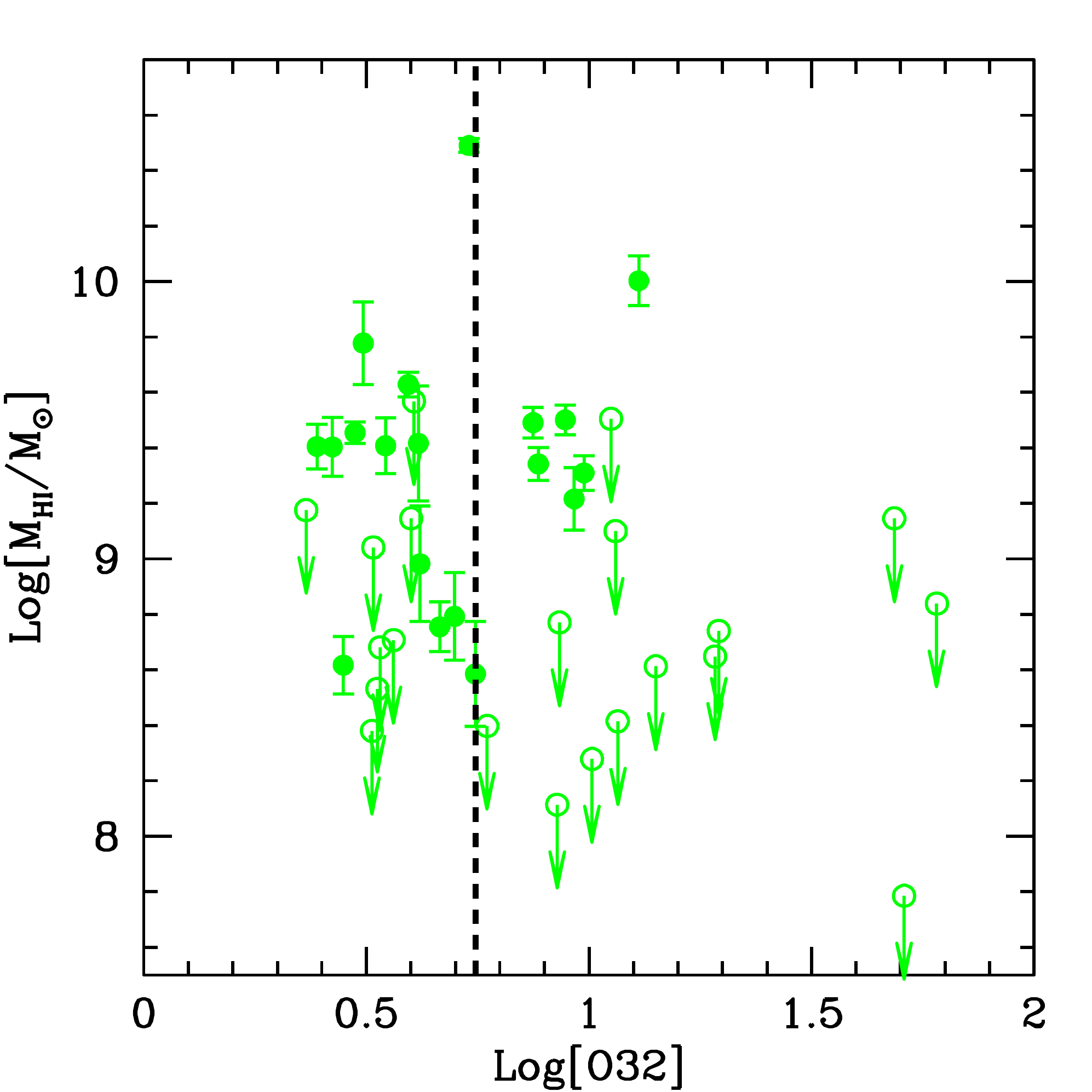}
\caption{The \hi\ mass of the Green Peas plotted against [A]~their absolute B-magnitude, $\mb$, and 
[B]~their O32 value. In [A], the dashed line indicates the $\mhi - \mb$ relation in the local Universe,
while the dotted lines indicate the $\pm 0.6$~dex (i.e. $\pm 2\sigma$) spread around the relation 
\citep{denes14}. A number of the Green Peas are seen to have \hi\ masses $\gtrsim +0.6$~dex above 
the local relation, while a few have \hi\ masses $\gtrsim 0.6$~dex below the relation. In [B], the dashed vertical line indicates the median O32 value, $\approx 5.5$.
\label{fig:o32}}
\end{figure}

Fig.~\ref{fig:o32}[A] plots the \hi\ mass of the Green Peas against their absolute B-magnitude $\mb$;
the dashed line indicates the $\mhi - \mb$ relation of galaxies in the local Universe, with the dotted 
lines indicating the $\pm 0.6$~dex ($\approx 2\sigma$) spread around the local relation 
\citep[e.g.][]{denes14}. While the majority of the Green Peas are seen to lie within the spread of the 
local $\mhi - \mb$ relation, it is interesting that nine of the 40 galaxies of our sample 
(i.e. $\approx 22$\%) lie $\gtrsim 0.6$~dex above it. This suggests that a significant fraction of 
Green Peas are gas-rich for their optical luminosity, possibly due to recent gas accretion from the 
circumgalactic medium or via a minor merger, or due to a gas-rich companion galaxy within the relatively 
large GBT or Arecibo beam. 

Conversely, five of the non-detections and two of the detections of \hii\ emission lie $\gtrsim 0.6$~dex 
{\it below} the local $\mhi - \mb$ relation. Further, most of the detections of \hii\ emission lie above 
the local relation, while most of the non-detections lie below it. This may suggest bimodality in the \hi\ 
properties of Green Pea galaxies, with one group having exhausted its neutral gas in the starburst (which 
may have been itself triggered by a recent gas acquisition via infall or a merger), and the other having 
only consumed a fraction of its neutral gas in the starburst. We note that a caveat to the above result is 
that the $\mhi - \mb$ relation of \citet{denes14} is based on an \hi-selected galaxy sample from the 
all-sky HIPASS survey \citep{zwaan05b}, and thus may be biased towards \hi-rich galaxies. As such, 
objects lying below the $\mhi - \mb$ relation of \citet{denes14} may not necessarily be \hi-poor galaxies.

Despite the above caveat, it is tempting to identify the first group of galaxies above with the objects 
that are likely to show leakage of Lyman-$\alpha$ and Lyman-continuum radiation (i.e.  to show Lyman-$\alpha$ 
{\it emission}). Eight of the Green Peas of our sample have Lyman-$\alpha$ spectroscopy, with seven detections 
of Lyman-$\alpha$ emission and one (J1448-0110) showing net Lyman-$\alpha$ absorption \citep{mckinney19}. 
Interestingly, five of the detections of Lyman-$\alpha$ emission are not detected 
in \hii\ emission, as expected from the above argument. However, two of the Lyman-$\alpha$-emitting galaxies, 
J0213+0056 and J1200+2719, do show detections of \hii\ emission, and with relatively high \hi\ masses, 
$\approx 3.2 \times 10^9 \ \msun$ (J0213+0056) and $\approx 1 \times 10^{10} \ \msun$ (J1200+2719). Further, 
both these galaxies are ``gas-rich'' systems in Fig.~\ref{fig:o32}[A]. \hii\ mapping studies are needed to test whether the Green Pea galaxy itself is \hi-rich, or if it might have a gas-rich companion. 
Such \hii\ mapping studies are also critical to directly determine the \hi\ column density 
distribution within the Green Peas, to test for the presence of \hi\ holes through which the Lyman-$\alpha$
and Lyman-continuum photons might escape. At any event, at the present time, no clear trend is apparent 
between the gas richness of the above 8 Green Peas and their Lyman-$\alpha$ escape fraction, with high 
Lyman-$\alpha$ escape fractions obtained at both high and low \hi\ masses (and gas richness) in the relatively 
small current sample \citep{mckinney19}. Deeper \hii\ emission studies would be needed to test the possibility 
of bimodality in the gas content of Green Pea galaxies.


We also examined the dependence of the \hi\ mass, \hi-to-stellar mass ratio, and \hi\ depletion time, 
on the metallicity (12+[O/H]) of the Green Peas of our sample, finding no evidence of a dependence 
of any of these properties on the metallicity. 

\citet{jaskot13} argued that the high luminosity ratio O32~$\equiv $~[O{\sc iii}]$\lambda$5007/[O{\sc ii}$\lambda$3727 
observed in a number of Green Pea galaxies at $z \approx 0.1-0.3$ makes them excellent candidates for the 
escape of ionizing Lyman-continuum radiation. A high Lyman-continuum leakage was indeed later found in 
galaxies with high O32 values, both at high redshifts \citep[e.g.][]{nakajima14,nakajima16,fletcher19} and 
low redshifts \citep[including Green Pea galaxies; e.g.][]{izotov16,izotov18a,izotov18b,izotov20}, for 
typical O32 values $\gtrsim 10$. One would expect easier leakage of Lyman-continuum photons from galaxies
with a lower average \hi\ column density, and also with a lower \hi\ mass. We hence examined the \hi\ properties 
in our Green Peas as a function of their O32 value; Fig.~\ref{fig:o32}[B] shows the measured \hi\ mass for the 
40 Green Peas plotted against the O32 values; the median O32 value is $\approx 5.5$, indicated by the dashed 
vertical line. Among the 20 Green Peas with O32 values below the median, there are 12 detections of \hii\ 
emission, with an average \hi\ mass of $5.6 \times 10^{9} \, \msun$, while for Green Peas with O32 values above 
the median there are 7 detections and an average \hi\ mass of $3.2 \times 10^{9} \, \msun$. Further, there is 
only a single detection of \hii\ emission in the 11 Green Peas with O32~$\geq 10$ (i.e. a detection fraction
of $0.091^{+0.21}_{-0.02}$), and 18 detections in the 29 Green Peas with O32~$< 10$ (i.e. a detection 
fraction of $0.62^{+0.18}_{-0.14}$). Thus, although the numbers are still small, both the detection rate 
and the average \hi\ mass appear to be significantly lower in galaxies with O32~$\gtrsim 10$,
consistent with the expected high Lyman-continuum leakage.

\citet{tilvi09} modelled star-formation in \lya\ emitters by assuming that the accretion of gas rapidly
results in star-formation, to obtain a star-formation efficiency ($f_\star$) of $\approx 2.5$\%. This is 
similar to the estimate of $f_\star \approx 4-8$\% obtained by \citet{baldry08}, by comparing the cosmic stellar 
mass density to the cosmic baryon density \citep[see also][]{fukugita98}. Assuming $f_\star \approx 2.5$\% yields a 
median star-formation timescale of $\tau_{\rm SF} \approx f_\star \times (\mhi/{\rm SFR})_{med} \equiv f_\star \times \tau_{\rm dep,med} \approx 15$~Myr. 
Interestingly, this is similar to the age of the young stellar population that dominates the starlight of the 
Green Peas of our sample \citep[$\approx 3-8$~Myr, with a median age of $\approx 4$~Myr; ][]{jiang19}. We note, however, that the above $f_\star$ estimates \citep{baldry08,tilvi09} are for all baryonic material, including the 
ionized gas. The agreement between the star-formation timescale and the age of the young stellar 
population in Green Peas might then suggest that the timescale of conversion from ionized gas to 
neutral gas is short in these galaxies.

\section{Summary}
\label{sec:summary} 

We report an Arecibo and GBT search for \hii\ emission from a large sample of Green Pea galaxies
at $z \approx 0.02-0.1$, obtaining 19~detections of \hii\ emission and 21 upper limits to the \hi\ mass, 
and yielding the first estimates of the gas content of these starbursting systems. The \hi\ properties of 
the majority of the Green Peas appear similar to those of galaxies in the local Universe, in terms of 
the \hi-to-stellar mass ratio and the $\mhi - \mb$ relations. However, a significant fraction of the Green Peas 
($\approx 22$\%) have an \hi\ mass that is $\gtrsim +0.6$~dex (i.e. $\gtrsim 2\sigma$) above the local 
$\mhi - \mb$ relation, indicating either recent gas accretion or a gas-rich companion galaxy. A similar fraction 
lie $\gtrsim 0.6$~dex below the local relation, suggesting possible bimodality in the gas properties of Green Peas. 
This large fraction of outliers ($\approx 30$\%) from the $\rm M_{HI} - M_B$ relation and the young ages of the 
stellar populations are indicative of a possible ``boom and bust'' nature of star-formation in Green Peas. 
Further, the \hi\ depletion times in Green Peas are an order of magnitude lower than values in local galaxies, 
indicating that the starburst activity will consume their \hi\ on timescales less than a Gyr. The detection rate of 
\hii\ emission appears low in galaxies with the highest O32 values, O32~$\geq 10$, consistent with the 
high Lyman-continuum leakage expected from these galaxies.

\acknowledgements
NK acknowledges support from the Department of Science and Technology via a Swarnajayanti Fellowship (DST/SJF/PSA-01/2012-13). This work was supported by the Department of Atomic Energy, under project 12-R\&D-TFR-5.02-0700. 
The Arecibo Observatory is a facility of the National Science Foundation operated under cooperative agreement by the University of Central Florida and in alliance with Universidad Ana G. Mendez, and Yang Enterprises,Inc.
The Green Bank Observatory is a facility of the National Science Foundation operated under cooperative agreement by Associated Universities, Inc.

\bibliographystyle{aasjournal}

\end{document}